\documentclass[twocolumn,prx,superscriptaddress,floatfix]{revtex4-1}
\usepackage{booktabs}
\usepackage{amsmath,amssymb,mathrsfs,amsthm}
\usepackage{natbib}
\usepackage{tabularx}
\usepackage{amsfonts}
\usepackage{subcaption}
\usepackage{amsmath}
\usepackage{comment}
\usepackage{bbold} 
\usepackage{hhline}
\usepackage{braket}
\usepackage{txfonts}
\usepackage{balance}
\usepackage{physics}
\usepackage{graphicx}
\usepackage{dcolumn}
\usepackage{multirow}
\usepackage{bm}
\usepackage[utf8]{inputenc}
\usepackage[english]{babel}
\usepackage[T1]{fontenc}
\usepackage{mathtools}
\usepackage[dvipsnames]{xcolor}
\usepackage{quantikz}
\usepackage[unicode=true,bookmarks=true,bookmarksnumbered=false,bookmarksopen=false,breaklinks=false,pdfborder={0 0 1},backref=false,colorlinks=true,citecolor=blue,linkcolor=blue]{hyperref}

\def\be{\begin{equation}}
\def\ee{\end{equation}}
\def\bea{\begin{eqnarray}}
\def\eea{\end{eqnarray}}

\newcommand{\RNum}[1]{\uppercase\expandafter{\romannumeral #1\relax}}

\newcommand{\gr}[1]{{\leavevmode\color{OliveGreen}#1}}
\newcommand{\re}[1]{{\leavevmode\color{red}#1}}
\newcommand{\floor}[1]{\lfloor {#1} \rfloor}

\begin{document}

\title{Qudit-based scalable quantum algorithm for solving the integer programming problem}

\author{Kapil Goswami}
\email{kapil.goswami@uni-hamburg.de}
\affiliation{Zentrum f\"ur Optische Quantentechnologien, Universit\"at Hamburg, Luruper Chaussee 149, 22761 Hamburg, Germany}

\author{Peter Schmelcher}
\email{peter.schmelcher@uni-hamburg.de}
\affiliation{Zentrum f\"ur Optische Quantentechnologien, Universit\"at Hamburg, Luruper Chaussee 149, 22761 Hamburg, Germany}

\author{Rick Mukherjee}
\email{rick-mukherjee@utc.edu}
\affiliation{Department of Physics \& Astronomy, University of Tennessee, Chattanooga, TN 37403, USA}
\affiliation{UTC Quantum Center, University of Tennessee, Chattanooga, TN 37403, USA}

\date{\today}

\begin{abstract}

Integer programming (IP) is an NP-hard combinatorial optimization problem that is widely used to represent a diverse set of real-world problems spanning multiple fields, such as finance, engineering, logistics, and operations research. It is a hard problem to solve using classical algorithms, as its complexity increases exponentially with problem size. Most quantum algorithms for solving IP are highly resource inefficient because they encode integers into qubits. In \cite{goswami2024integer}, the issue of resource inefficiency was addressed by mapping integer variables to qudits. However, \cite{goswami2024integer} has limited practical value due to a lack of scalability to multiple qudits to encode larger problems. In this work, by extending upon the ideas of \cite{goswami2024integer}, a circuit-based scalable quantum algorithm is presented using multiple interacting qudits for which we show a quantum speed-up for solving the integer programming problem. The quantum algorithm consists of a \textit{distillation function} that efficiently separates the feasible from the infeasible regions, a phase-amplitude encoding for the cost function, and a quantum phase estimation coupled with a multi-controlled single-qubit rotation for optimization. We prove that the optimal solution has the maximum probability of being measured in our algorithm. The time complexity for the quantum algorithm is shown to be $O\left(r_P\cdot \big(m\cdot n^2\cdot \log d + \frac{d^{n/2}}{\sqrt{N_{\mathbf{y}_s}}} + n \cdot \epsilon_{\text{QPE}}^{-1}\big)\right)$ for a problem with the number of variables $n$ taking $d$ integer values, satisfying $m$ constraints with a precision of $\epsilon_{\text{QPE}}$ and $r_P$ repetitions required for reaching a success probability $P$. Compared to the classical time complexity of brute force $O(d^n)$ and the best classical exact algorithm $O((\log{n})^{O(n)})$, it incurs a reduction of $\sim d^{n/2}$ in the time complexity (a quadratic speed-up for polynomial number of repetitions and precision requirements) in terms of $n$ for solving a general polynomial IP problem.
\end{abstract}

\maketitle

\section{Introduction}
\label{sec:introduction}

One pertinent question while building new quantum algorithms is whether they can provide a quantum advantage over their classical counterparts \cite{dalzell2023quantum,montanaro2016quantum,cerezo2021variational,shor2002introduction}.
Another related question is to look for problems for which the quantum algorithms can play a role in solving the NP-hard problems efficiently \cite{mandal2023review,abbas2024challenges}.
In this work, both of these questions are addressed for a well-known combinatorial optimization problem with widespread real-world applications: the integer programming (IP) problem \cite{wolsey2020integer,schrijver1998theory}.

Quantum advantage between algorithms can be assessed by comparing the number of queries made to an \textit{oracle} by each algorithm to find the solution.
An oracle is a ``black-box" subroutine that computes a function $f$ in a single step, with access being limited to the input and output only.
This led to the development of many oracle-based quantum algorithms, such as the Deutsch-Jozsa (DJ), Bernstein-Vazirani (BV), and Simon's algorithms \cite{deutsch1992rapid,deutsch1985quantum,bernstein1993quantum,nagata2020generalization,simon1997power}, where quantum advantage can be explicitly evaluated compared to the classical algorithms.
Since practical realization of oracles using quantum circuits can be difficult, the gate-based approaches provide alternatives for the oracle subroutines, some of them are Shor's algorithm, the Quantum Fourier Transform (QFT), quantum phase estimation (QPE), and Grover's algorithm \cite{shor1994algorithms, coppersmith2002approximate, kitaev1995quantum, grover1997quantum}.
Although implementing oracles with gates can be cumbersome and expensive, nevertheless, these gate-based approaches set tangible targets required for quantum advantage, such as the number of qubits, the types of gates, the number of gates, and the acceptable error rates, all of which contribute to the threshold arguments \cite{knill2005quantum}.

We employ a similar approach to develop a quantum algorithm for integer programming, specifically utilizing a gate-based algorithm that incorporates qudits. Recently, there have been many developments in quantum computing using qudits instead of qubits \cite{wang2020qudits,Chi2022,Deller2023,Nikolaeva2024,Kim2024,Pudda2024}. Qudits are $d$-dimensional objects \cite{wang2020qudits} that provide a larger Hilbert space ($d^n$ for $n$ $d$-dimensional qudits) to explore and store information. This helps to reduce circuit complexity and enhances the efficiency of the computations performed \cite{wang2020qudits}. In Ref.~\cite{goswami2024integer}, small IP problems were solved using a single \textit{qudit}, which was a significant resource advantage compared to other quantum algorithms for IP. It, however, lacked two crucial aspects: an explicit quantum advantage and scalability to larger problem sizes.
In this work, a scalable quantum algorithm is proposed that utilizes qudits to solve IP with a quadratic speed-up for polynomial number of repetitions and precision requirements in time complexity compared to classical exhaustive search.

We propose to use a qudit array to store the integer variables, and a qubit array keeps track of the constraint satisfiability; both of the arrays are entangled in the circuit.
Solving a given IP problem is divided into two tasks, similar to Ref.~\cite{goswami2024integer}: satisfying the constraints and optimizing the cost function. The constraints are satisfied by defining a \textit{distillation function} that separates the feasible from the infeasible region; the qudit array is made to collapse onto the subspace where constraints of the IP problem are satisfied by measuring the qubit array in a specific state. This is followed by applying well-established methods such as quantum phase estimation and multi-controlled rotation, and the probability of the optimal solution is maximized. Apart from resource efficiency over qubits, our proposed algorithm shows reduced time complexity over classical exhaustive search algorithm. For a bounded-polynomial IP problem defined with $n$ integers taking $d$ distinct values and $m$ constraints, the time for the quantum algorithm scales as $T_Q\sim O(r_P(d^{n/2} + m\cdot n^2\cdot \log{d} + n/\epsilon_{\text{QPE}}))$ (in leading order terms) to solve the problem within $\epsilon_{\text{QPE}}$ precision and requiring $r_P$ independent repetitions to achieve a success probability $P$. This can provide a significant reduction from the classical brute force time complexity of $T_{Cl}\sim O(d^n)$, which provides a general baseline for bounded polynomial-IP, including non-convex nonlinear instances. Along with significantly reduced time complexity, under ideal operations and exact cost encoding, the optimal solution has the largest probability of measurement in the qudit-qubit space.

Several quantum optimization methods utilize amplitude amplification or Grover search, including the minimum-finding algorithm of D\"urr and H{\o}yer \cite{durr1996minimum}, the Grover adaptive search framework of Baritompa, Bulger, and Wood \cite{baritompa2005grover}, and the construction of Gilliam, Woerner, and Gonciulea for constrained polynomial binary optimization \cite{gilliam2021grover}. For example, D\"urr--H{\o}yer's work finds the minimum from $N$ entries using $O(\sqrt{N})$ oracle queries at constant success probability by repeatedly searching for optimal entries smaller than the best one so far. Although this framework may be applied to bounded integer programming using suitable feasibility and objective oracles, the original work neither constructs these oracles nor presents any encoding scheme to address the integer programming problem. The role of amplitude amplification also differs between these approaches. In D\"urr--H{\o}yer minimum finding and Grover adaptive search, amplitude amplification searches for the optimum by selecting candidates whose objective values lie below the current best value \cite{durr1996minimum,baritompa2005grover,gilliam2021grover}. In our algorithm, amplitude amplification is instead used to enhance the probability of the feasible region, independently of its objective value. The objective is processed through cost-phase encoding, quantum phase estimation, and controlled rotations. Thus, our construction separates feasible-state preparation from objective-dependent amplitude processing, without an adaptive objective-threshold search.

Here is the outline of the manuscript. In Section~\ref{PD}, we briefly define a general IP problem. Section~\ref{Alg} covers the main techniques that are used to construct the quantum algorithm. The complexity analysis and the success probability for the algorithm are described in Section~\ref{COMPLEXAN}. In Section~\ref{results}, we showcase the numerical implementation of the quantum algorithm along with the comparison with classical algorithms. 
We provide our conclusions in Section~\ref{conc}, where we also make comments about the robustness of the algorithm on real noisy hardware.

\section{Problem definition \label{PD}}
The classical algorithms for many optimization problems are reaching their limits nowadays, presenting both new opportunities and challenges for quantum algorithms \cite{mandal2023review,abbas2024challenges}.
The optimization problems often fall into the NP-hard complexity class, making it exponentially difficult to solve using classical algorithms as the problem size increases \cite{karp1972complexity,kannan1978computational,papadimitriou1982complexity}.
It is widely believed that quantum algorithms, much like their classical counterparts, cannot solve NP-hard problems in polynomial time \cite{bennett1997strengths,aaronson2005complexity}. However, they may still offer significant advantages, such as a polynomial speed-up in complexity, resulting in a huge practical advantage \cite{grover1997quantum,paschos2009overview,pirnay2024principle}.
There are, broadly speaking, two mathematical structures that describe these optimization problems, namely quadratic unconstrained binary optimization (QUBO) and integer programming (IP) \cite{wolsey2020integer,schrijver1998theory}. In the context of quantum algorithms, QUBO is a preferred choice due to its natural mapping to spin Hamiltonians \cite{lucas_ising_2014,barahona1988application,glover_tutorial_2019,farhi2014quantum,peruzzo_variational_2014,goswami2023solving,ohzeki_quantum_2010, hadfield_quantum_2019}. For classical algorithms, most optimization problems are solved using integer programming (IP) or, in the general case, mixed-integer programming (MIP) rather than QUBO \cite{pochet2006production,magatao2004mixed,kannan1978computational}. This is because IP offers a more general mathematical structure for encoding real-world problems, which often contain discrete variables and multiple constraints \cite{wolsey2020integer,papadimitriou1982complexity}.
The quantum algorithms for solving IP are highly resource-inefficient due to the lack of a direct mapping to encode discrete-integer variables onto a quantum system; instead, they rely on mapping the integer variables to binary \cite{chang2020hybrid,okada2019efficient,svensson2023hybrid}. One demonstration of mapping an optimization problem directly to a quantum system is presented in Ref.~\cite{TSP} that utilizes a geometric interpretation of a qubit.

For a general bounded variables IP problem with $n$ decision variables each taking $d$ values, the goal is to maximize the polynomial-cost function $C(\mathbf{x})$ represented as
    \begin{equation}
        C(\mathbf{x}) = c_1 x_{1}\cdots x_{n'} + \cdots + c_{v} x_{1} \cdots x_{n''},
    \end{equation}
where $n', n'' \in \mathbb{Z}^+ \leq n$, $v$ is the number of monomial terms in the cost function, $c_{j}$ are the corresponding real-valued coefficients, and $x_i \in \{0,...,d-1\}$. There are $ m$ constraints for the problem given as
\begin{align}
\begin{aligned}
C_1 &= a_{11} x_{q_1} \cdots x_{q_{k_1}} + \cdots \leq h_1, \\
&\vdots \\
C_m &= a_{m1} x_{r_1} \cdots x_{r_{k_m}} + \cdots \geq h_m,
\end{aligned}
\end{align}
where $q_j,r_j \in \mathbb{Z}^+$, $a_{uv}$ are the real-valued coefficients for each term in the constraint polynomials, and $C_i$ are known finite-degree polynomial functions. The constraints $C_i$ and the cost function $C(\mathbf{x})$ can contain terms of any order, e.g., linear, quadratic, cubic, and so on, in the decision variables. All IP problems can be reformulated to satisfy the following conditions using a constant shift to the variables.
\begin{itemize}
    \item The inequality in the constraint ($C_i$) is strictly \textit{less} than a certain integer ($h_i$): $C_i < h_i$.
    \item The variables $\mathbf{x} \in \mathbf{Z^+} \cup \{0\}$ in the constraint take \textit{positive} and zero integer values.
\end{itemize}
This form is suitable for implementing the qudit-based quantum algorithm.

\begin{figure*}[tp]
    \centering
    \includegraphics[width = 1\linewidth]{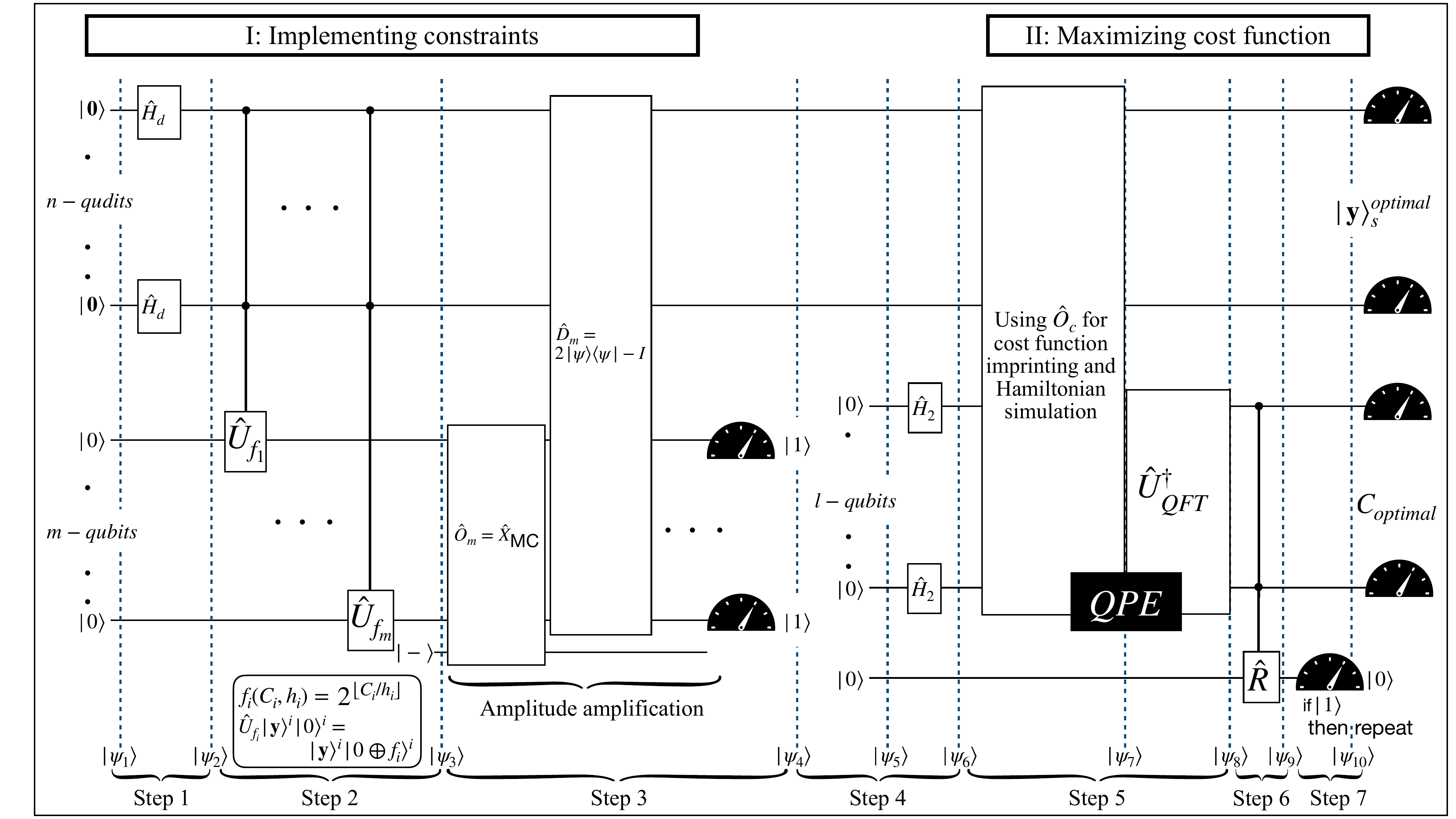}
    \caption{Quantum circuit of the qudit-based algorithm for solving a general IP problem with $n$ variables and $m$ constraints. It is divided into two stages: implementing constraints and maximizing the cost function. The $n$-qudit register stores the integer variables, and the $m$-qubit register indicates the satisfiability of the constraints. The subroutines for the constraints satisfying stage include $n$-Hadamard gates $\hat{H}_d$ for qudits, $m$ one-sparse entangling unitary operators $\hat{U}_{f_i}$ in sequence, $\sim d^{n/2}$ times application of the Grover operator $\hat{O}_m\cdot \hat{D}_m $ for amplitude amplification, and $m$-measurements for the qubit register. The next part, namely, maximizing the cost function, begins with a collapsed wave function of the $n$-qudit superposition state, a qubit register with $ l$ qubits, and a single qubit. The subroutines for this part are: $l$-Hadamard gates $\hat{H}_2$ for qubits, cost function imprinting using a QPE procedure by Hamiltonian simulation of $\hat{O}_c$ (see Step 5 in the main text for the structure of $\hat{O}_c$), a multi-qubit controlled rotation gate $\hat{R}$, and $l+1$-qubit plus $n$-qudit measurements. The measurements at the end are probabilistic, where the problem is solved with the highest probability compared to other possibilities.}
    \label{Dalg}
\end{figure*}

\section{Qudit-based quantum algorithm \label{Alg}}

A brief description of the qudit-based quantum algorithm for solving the IP problem, as shown by the circuit diagram in Fig.~\ref{Dalg}, consisting of two stages, is provided here, and the detailing of the steps follows.

Our quantum algorithm for integer programming encodes each decision variable directly into a qudit (a multi-level quantum system), so the discrete multi-level system naturally maps to the integer structure of the problem.
For representing an IP with $ n$ variables taking $d$-values, $n$ qudits each of dimensions $d$ are used, while $ m$ qubits represent $m$ constraints.
The algorithm begins by preparing a uniform superposition over all the possible qudit states using an array of $n$-Hadamard gates to $n$-qudits while the qubits are left unchanged. The goal is to find a subspace of qudit states satisfying all the constraints of the problem, termed as the \textit{feasible region}.
This is done by applying a feasibility distillation function that filters out constraint-violating states.
The unitary operator corresponding to the distillation function flips the qubit conditioned on the state of the qudit register, which entangles the many-body qudit states satisfying all the constraints simultaneously to a unique multi-qubit state. A Grover-like amplitude amplification is then applied to the multi-qubit state, and its measurement results in a collapsed subspace of qudit states forming the feasible region. This procedure of the algorithm is termed the constraint-satisfying stage, as illustrated in Fig.~\ref{Dalg}.

The cost function is encoded into the quantum state through controlled phase and amplitude mappings, defining the cost function optimization stage. In this stage, the measured constraint register is discarded, and a new qubit register is introduced to store the value of the cost function.
This qubit register stores the cost function of the IP corresponding to the qudit-states by performing a QPE procedure. The intermediate states of the new qubit register during the QPE procedure are entangled with the feasible states of the qudit register, accumulating the phase corresponding to the cost function value. A single ancillary qubit is then rotated, with a control on the qubit register, amplifying the probability of the optimal solution so that a measurement yields the best feasible integer assignment with high probability.

The time complexity scaling of this algorithm is the best known for a general IP problem, for which a proof is provided later.
This work is highly relevant as it goes beyond qubit-only optimization approaches (such as the quantum approximate optimization algorithm, QAOA, or the variational quantum eigensolver, VQE) by directly exploiting qudit encoding, which is increasingly accessible in modern trapped-ion and photonic quantum hardware. Unlike heuristic quantum methods, this algorithm is exact for integer programming and achieves improvements over classical exponential-time algorithms.

\subsection{Stage I: Implementation of the constraints}
\label{stage1}

\noindent \textit{Step 1. Initialization.}
The system is initialized with a $d$-dimensional qudit register consisting of $ n$ qudits representing $n$ integer variables each taking values in $[0,d-1]$.
A second register with $m$ qubits is introduced, corresponding to $m$ constraints in the problem, where the state $\ket{1}$ or $\ket{0}$ of each qubit determines if a constraint ($C_i<h_i$) is satisfied or not, respectively. The registers are initialized in the state $\ket{\mathbf{0}}$ for qudits and $\ket{0}$ for qubits, where states in bold denote qudits.
Thus, the state of the system with the two registers shown as $\ket{\psi_1}$ in Fig.~\ref{Dalg} is,
 \begin{equation}
\begin{split}
\begin{aligned}
  \ket{\psi_1} = \ket{\mathbf{0}}^{\otimes n} \ket{0}^{\otimes m},
    \end{aligned}
\end{split}
\label{1}
\end{equation}
where the tensor product $\otimes$ in the state space is implicit.

One advantage of using qudits is that they offer a large Hilbert space $d^n$ where each multi-qudit state naturally encodes the corresponding integer variable assignments. Applying $n$-Hadamard gates $\hat{H}_d$ with dimension $d$ only on the qudit register results in a uniform superposition of all the possible multi-qudit states, while leaving the qubit register unchanged. The explicit form of the $d$-dimensional Hadamard gate is provided in Appendix~\ref {Hadamards}.
The state of the system is,
       \begin{equation}
\begin{split}
\begin{aligned}
\ket{\psi_2} &= \qty(\hat{H}_d^{\otimes n}\ket{\mathbf{0}}^{\otimes n}) \ket{0}^{\otimes m}
   = \qty(\frac{1}{d^{n/2}}\sum_{y=0}^{d^n-1}\ket{\mathbf{y}}) \ket{0}^{\otimes m},
    \end{aligned}
\end{split}
\label{2}
\end{equation}
where $\ket{\mathbf{y}}$ is a shorthand notation for the multi-qudit register defined as,
\begin{equation}
\begin{split}
\begin{aligned}
 \ket{\mathbf{y}} &\equiv \ket{\mathbf{x_1,...,x_b,...,x_n}} = \ket{\mathbf{x_1}} \otimes ... \otimes \ket{\mathbf{x_b}} \otimes ... \otimes \ket{\mathbf{x_n}} \\
 &\quad \quad \mathbf{x_b} \in \{0,1,...,d-1\} \text{ }\text{ } \forall \mathbf{b} \in [1,n],
    \end{aligned}
\end{split}
\label{2_1}
\end{equation}
where $\mathbf{y}\in [0,d^n-1]$ and each $\ket{\mathbf{y}}$ represents a multi-qudit configuration. As an example, for $n=2$ and $d=3$, $\mathbf{y} \in [0,8]$, where the mapping is such that $\ket{\mathbf{y}=0} \mapsto \ket{00}$, $\ket{\mathbf{y}=1} \mapsto \ket{01}$, $\cdots$, $\ket{\mathbf{y}=8} \mapsto \ket{22}$.
The total number of Hadamard gates for the sub-routine $\ket{\psi_1} \mapsto \ket{\psi_2}$ is equal to the number of qudits, i.e., $\sim n$.

\noindent \textit{Step 2. Implementation of the constraint-satisfying distillation function.}
One of the main bottlenecks in solving the IP is finding the variable assignments that satisfy all the constraints simultaneously. This is a tedious task as the total space of the variables is exponentially large ($d^n$), out of which a small subspace needs to be extracted, contributing to the IP problem being NP-hard.
The goal is to find a subspace of qudit states satisfying all the constraints of the problem, termed as the \textit{feasible region}.
The qubit register plays a crucial role in this procedure as the state $\ket{0}/\ket{1}$ of each qubit determines the corresponding constraint's satisfiability. To extract the constraint information from the qudit register and store it in the qubit register, we define a distillation function $f$ and use it to entangle the two registers.
The function $f_i$ corresponding to a constraint $C_i<h_i$ is devised as follows,
\begin{equation}
\begin{split}
f_i(C_i,h_i) &= 2^{\floor{\frac{C_i}{h_i}}} \\
\implies  f_i(C_i,h_i) &=  \begin{cases}
      2^{\floor{\frac{C_i}{h_i}}} = \text{even } \mathbf{Z^+} & : \text{Infeasible region}  \\
      2^{\floor{\frac{C_i}{h_i}}} = 1 & : \text{Feasible region},
    \end{cases}
\end{split}
\label{fun}
\end{equation}
where $\floor{C_i/h_i}$ returns the nearest integer value that is less than $C_i/h_i$. If the constraint $C_i < h_i$ is satisfied, the value of $f_i$ is $1$; otherwise, it's an even integer. The details leading to the construction of the function $f_i$ are provided in Appendix~\ref {fdis}.
An operator definition of the function $f_i$ is required to implement it in a circuit-based algorithm.

The construction of the operator $\hat{U}_{f_i}$ implementing $f_i$ is such that it acts on a restricted subspace of the full qudit Hilbert space. Specifically, it operates on the subset of basis states \(\ket{\mathbf{y}}^i\), where only the qudits corresponding to the variables that influence the value of \(C_i\) are allowed to change. All other qudits are held fixed. This ensures that \(\hat{U}_{f_i}\) modifies only the relevant degrees of freedom needed to evaluate the constraint.
The operator \(\hat{U}_{f_i}\) divides the set of multi-qudit states \(\ket{\mathbf{y}}^i\) into two disjoint subsets;
\(\ket{\mathbf{y}}_s^i\), consisting of states that satisfy the constraint \(C_i < h_i\), and \(\ket{\mathbf{y}}_k^i\), consisting of those that do not.
$\hat{U}_{f_i}$ entangles all the constraint-satisfying multi-qudit states $\ket{\mathbf{y}}_s^i$ with the qubit state $\ket{1}^i$, and all the other states $\ket{\mathbf{y}}_k^i (\neq \ket{\mathbf{y}}_s^i)$ with the qubit state $\ket{0}^i$.
Thus, the unitary operator $\hat{U}_{f_i}$ corresponding to the function $f_i$ applied to the set of states $\ket{\mathbf{y}}^i \ket{0}^i$ acts as,

\begin{widetext}
\begin{equation}
\begin{split}
\begin{aligned}
     \hat{U}_{f_i} \ket{\mathbf{y},0}^i &=  \ket{\mathbf{y},0 \oplus_2 2^{\floor{\frac{C_i}{h_i}}}}^i
    = \begin{cases}
      \ket{\mathbf{y},0}^i_{k} & : \text{Infeasible region} \\
       \ket{\mathbf{y},1}^i_{s} & : \text{Feasible region}.
    \end{cases}
    \end{aligned}
\end{split}
\label{equ31}
\end{equation}
\end{widetext}
where $\oplus_2$ performs Boolean addition.

Since the input and output states corresponding to the unitary $\hat{U}_{f_i}$ are known, the corresponding matrix can be constructed. The unitary $\hat{U}_{f_i}$ turns out to be a 1-sparse matrix; i.e., the matrix consists of only one non-zero element in each row (1-sparse). The 1-sparsity of the matrix is a direct result of Eq.~\eqref{equ31}, as the operator maps the input state either to itself or permutes it with another state where the qubit has the opposite spin. In general, any operator acting on $q$ qubits can be expanded in the basis of Pauli strings, $\hat{A}=\sum_{\hat{P}\in\{\hat{I},\hat{X},\hat{Y},\hat{Z}\}^{\otimes q}} a_{P}\,\hat{P}$~\cite{nielsen2010quantum} (and analogously in the generalized Weyl--Heisenberg basis for qudits~\cite{schwinger1960unitary,gottesman1999fault,wang2020qudits}). Sparse matrices can be implemented through such decompositions, or through sparse-access oracles and block encodings~\cite{berry2014exponential}. However, the number of non-vanishing terms, and the efficiency of the implementation, depends on the structure of the matrix and can grow exponentially with the system size in the worst case. For the 1-sparse unitaries considered here, the structure makes it a relatively simpler case. Since $\hat{U}_{f_i}$ is a permutation matrix which only exchanges $\ket{\mathbf{y},0}^i \leftrightarrow \ket{\mathbf{y},1}^i$ for feasible $\mathbf{y}$ (Eq.~\eqref{equ31}), it decomposes into local swaps (transpositions) between pairs of basis states that differ only in the state of the $i$-th qubit. In other words, $\hat{U}_{f_i}$ acts as a NOT gate on the $i$-th qubit, controlled by the Boolean function that evaluates the $i$-th constraint on the qudit register. Its cost is set by the reversible evaluation of this function $f_i(C_i,h_i)$, e.g., arithmetic and comparator circuits. A full gate-level synthesis of this hybrid qudit--qubit unitary is beyond the scope of this work, hence, to leading order, the gate count scales polynomially in $n$ and $\log d$, as described next.

In general, there are multiple strategies for implementing $\hat{U}_{f_i}$~\cite{camps2024explicit,gaidai2025decomposition,li2025nearly,gonzales2024arbitrary}. It is crucial to note that an arbitrary permutation of the $d^n$ basis states cannot, in general, be implemented with a polynomial number of gates, and efficiency relies on the structure of the permutation. In our case, the permutation is generated by an efficiently computable constraint function, and we use the permutation-matrix decomposition to break the unitary into gates, which for such structured permutations requires $O(n^2)$ qudit gates and can be optimized to use $O(n^2\log{d})$ qubit gates. An example of the circuit implementation for $\hat{U}_{f_i}$ is given in Appendix~\ref{fdis}. Furthermore, since $\hat{U}_{f_i}$ is a permutation matrix, it can be implemented exactly, i.e., without approximation error. After a sequential application of $m$ such $\hat{U}_{f_i}$ where $i\in[1,m]$ corresponding to $m$ constraints, the qudit states lying in the feasible region of the problem are entangled with the state $\ket{1}^{\otimes m}$ of the qubit register. Thus, the number of gates for $m$ 1-sparse unitary operators is $\sim O(m \cdot n^2\cdot\log{d})$

The resulting state after this sequential $\hat{U}_{f_i}$ is,
\begin{widetext}
\begin{equation}
\begin{split}
\begin{aligned}
\ket{\psi_3} &= \prod_{i=1}^{m}\hat{U}_{f_i} \ket{\psi_2}
   = \frac{1}{d^{n/2}}\Bigg (\sum_{(k_\gamma,\gamma)} \ket{\mathbf{y}}_{k_\gamma}\ket{q(\mathbf{y}_{k_\gamma})} + \sum_{\mathbf{y}_s\in\mathcal{S}} \ket{\mathbf{y}}_s\ket{ 1}^{\otimes m}\Bigg),
    \end{aligned}
\end{split}
\label{31}
\end{equation}
\end{widetext}

where $\ket{q(\mathbf{y}_{k_\gamma})}$ is the multi-qubit computational basis state associated with the assignment $\mathbf{y}_{k_\gamma}$, which satisfies exactly $\gamma\in\{0,\ldots,m-1\}$ constraints. Thus, the first summation runs over pairs $(k_\gamma,\gamma)$, where $k_\gamma$ enumerates assignments satisfying $\gamma$ constraints.

The constraint-register state $\ket{q(\mathbf{y})}$ associated with an assignment $\mathbf{y}$ is
\begin{equation}
\ket{q(\mathbf{y})}=\bigotimes_{i=1}^{m}\ket{b_i(\mathbf{y})},
\qquad
b_i(\mathbf{y})=
\begin{cases}
1,& C_i(\mathbf{y})<h_i,\\
0,& C_i(\mathbf{y})\geq h_i.
\end{cases}
\end{equation}
The Hamming weight of this bitstring is the number of satisfied constraints. Assignments with the same Hamming weight can have different bitstrings, which are retained through the dependence on $\mathbf{y}$.  The detailed steps for applying $\hat{U}_{f_i}$ sequentially to $\ket{\psi_2}$ and give $\ket{\psi_3}$ are provided in Appendix~\ref{Sequential}.

The state $\ket{\mathbf{y}}_s$ satisfies all the $m$ constraints $C_i$, while $\ket{\mathbf{y}}_{k_\gamma}$ for $\gamma\in\{0,\ldots,m-1\}$ are the states that contradict at least one of the constraints. These are individual computational basis states labelled by assignments in the sets
\begin{widetext}
\begin{equation}
\begin{split}
\mathcal{S}&=\left\{\mathbf{y}\in\{0,\ldots,d-1\}^n:
C_j(\mathbf{y})<h_j,\ \forall j\in\{1,\ldots,m\}\right\},\\
\mathcal{K}_{\gamma}&=\left\{\mathbf{y}\in\{0,\ldots,d-1\}^n:
\left|\{j:C_j(\mathbf{y})\geq h_j\}\right|=m-\gamma\right\}.
\end{split}
\label{sets}
\end{equation}
\end{widetext}
Here, $|\cdot|$ is the cardinality of the set, and $\mathcal{K}_{\gamma}$ contains assignments for which exactly $m-\gamma$ constraints are not satisfied. For example, for $\mathbf{y}\in\mathcal{K}_0$, $\ket{q(\mathbf{y})}=\ket{0\cdots 00}$; for $\mathbf{y}\in\mathcal{K}_1$, $\ket{q(\mathbf{y})}$ is one of $\ket{0\cdots 01},\ket{0\cdots 10},\dots,\ket{1\cdots 00}$; and for $\mathbf{y}\in\mathcal{K}_{m-1}$, it is one of $\ket{1\cdots 10},\ket{1\cdots 01},\dots,\ket{01\cdots 1}$, depending on which constraints $\mathbf{y}$ satisfies.

Interestingly, the entangled multi-qubit and multi-qudit registers lead to the separation of states with varying degrees of constraint satisfaction.
If the feasible region has no solution (infeasible problem), the system can focus on the states entangled with one less constraint to provide the best solution in that case.

\noindent \textit{Step 3. Reductions to the feasible subspace of IP.}
The state $\ket{\psi_3}$ (Eq.~\eqref{31}) with a separation of feasible and infeasible states needs to be measured such that the system collapses to the states in the feasible region $\ket{\mathbf{y}}_s$ in one step. Only the measurement of the qubit register needs to be performed such that all the qubits are in the state $\ket{1}$.
However, the amplitude of the state $\ket{1}^{\otimes m}$ corresponding to the feasible region $\ket{\mathbf{y}}_s$ may not be maximum. Hence, the goal is to amplify the amplitude of $\ket{1}^{\otimes m}$ by applying Grover's operator for the multi-qubit case.
The general procedure for Grover's amplitude amplification requires a combination of the phase flip operator $\hat{O}_m$ and the diffusion operator $\hat{D}_m$, and the detailed description is provided in Appendix~\ref{Grovers}.

The operator $\hat{O}_m$ applies a phase flip and marks the target state $\ket{1}^{\otimes m}$ with a $-1$ phase shift while the rest of the states remain unchanged.
\begin{equation}
\begin{split}
\begin{aligned}
\hat{O}_m &\ket{1}^{\otimes m} = -\ket{1}^{\otimes m} \\
\hat{O}_m &\ket{x} = \ket{x}, \quad \forall \ket{x} \neq \ket{1}^{\otimes m}.
    \end{aligned}
\end{split}
\label{O_m}
\end{equation}
For the implementation of $\hat{O}_m$ in a circuit, an ancilla qubit is introduced and is prepared in $\ket{-}$ state. A multi-controlled $X$ gate, represented as $\hat{X}_{\text MC}$ is applied to the $ m$-qubit register along with $\ket{-}$ which results in the state $(-1)\cdot\ket{1}^{\otimes m}\ket{-}$, marking the state $\ket{1}^{\otimes m}$ with a $-1$ phase shift. The application of the operator $\hat{O}_m$ to the state $\ket{\psi_3}$ is shown in Appendix~\ref{Grovers} and the resulting state as follows,
\begin{widetext}
\begin{equation}
\begin{split}
\begin{aligned}
\hat{O}_m\ket{\psi_3}\ket{-} &= \frac{1}{d^{n/2}}\hat{X}_{MC}\Bigg (\sum_{(k_\gamma,\gamma)} \ket{\mathbf{y}}_{k_\gamma}\ket{q(\mathbf{y}_{k_\gamma})} + \sum_{\mathbf{y}_s\in\mathcal{S}} \ket{ \mathbf{y}}_s\ket{ 1}^{\otimes m}\Bigg)\ket{-} \\
&= \frac{1}{d^{n/2}}\Bigg (\sum_{(k_\gamma,\gamma)} \ket{\mathbf{y}}_{k_\gamma}\ket{q(\mathbf{y}_{k_\gamma})} - \sum_{\mathbf{y}_s\in\mathcal{S}} \ket{\mathbf{ y}}_s\ket{ 1}^{\otimes m}\Bigg)\ket{-}.
    \end{aligned}
\end{split}
\label{O_m_apply}
\end{equation}
Once the target state $\ket{\mathbf{y}}_s\ket{ 1}^{\otimes m}$ is marked, the diffusion operator $\hat{D}_m$ is applied, which is used to reflect the state around the average; it is defined below and further elaborated in the Appendix~\ref{Grovers}.

In a circuit, the diffusion operator is implemented using a multi-controlled-Z gate
represented as $\hat{Z}_{\text{MC}}$, $n+m$ Pauli-X gates denoted by
$\hat{X}^{\otimes (n+m)}$, and the state-preparation unitary
$\hat{U}_{\psi} = \hat{U}_{f}\left(\hat{H}^{\otimes n}\otimes\hat{I}^{\otimes m}\right)$,
which generates $\ket{\psi_3} = \hat{U}_{\psi}\ket{0}^{\otimes n}\ket{0}^{\otimes m}$,
together with its inverse, in the following way,
\begin{equation}
\begin{split}
\hat{D}_m &= -\,\hat{U}_{f}\left(\hat{H}^{\otimes n}\otimes\hat{I}^{\otimes m}\right)
             \hat{X}^{\otimes (n+m)}\,\hat{Z}_{\text{MC}} \hat{X}^{\otimes (n+m)}
             \left(\hat{H}^{\dagger\,\otimes n}\otimes\hat{I}^{\otimes m}\right)\hat{U}_{f}^{\dagger} \\
          &= \hat{U}_{\psi}\left(2\left(\ket{0}\bra{0}\right)^{\otimes (n+m)} - \hat{I}\right)\hat{U}_{\psi}^{\dagger} = 2\ket{\psi_3}\bra{\psi_3} - \hat{I}.
\end{split}
\label{Dmm}
\end{equation}
\end{widetext}
where the operator $\hat{U}_{f}$ is constructed from the tensor product of all $\hat{U}_{f_i}$,
$\hat{U}_{f} = \prod_{i=1}^{m}\hat{U}_{f_i}$, and its inverse, $\hat{U}_{f}^{\dagger} = \prod_{i=m}^{1}\hat{U}_{f_i}^{\dagger}$, applies the operations in reverse order. $\hat{U}_{f}$ along with the Hadamard layer transforms the computational basis states such that the reflection is performed about the state $\ket{\psi_3}$, which is a non-uniform
superposition state. This differs from the standard diffusion operator, which is applied
to a uniform superposition state and uses only Hadamard gates.
The operator $\hat{D}_m$ operates on the complete entangled state $\ket{\psi_3}$ of the
$n$ qudits and $m$ ancilla qubits.
In this transformed basis, $\hat{U}_{\psi}^{\dagger}$ maps $\ket{\psi_3}$ onto
$\ket{0}^{\otimes (n+m)}$, so this state captures the information of the average amplitude
of the overall state. The $X$ gate flips each qubit and thereby maps
$\ket{0}^{\otimes (n+m)}$ to the target state $\ket{1}^{\otimes (n+m)}$. The
multi-controlled Z gate $\hat{Z}_{\text{MC}}$ performs the inversion around the mean by
applying a phase of $-1$ to $\ket{1}^{\otimes (n+m)}$, which is the image of
$\ket{0}^{\otimes (n+m)}$ after the $\hat{X}^{\otimes (n+m)}$ operation. The last
operations, $\hat{X}^{\otimes (n+m)}$ followed by $\hat{U}_{\psi}$ (i.e.\ the Hadamard
layer and then $\hat{U}_{f}$), return the system to the original basis. Since
$\hat{X}^{\otimes (n+m)}\hat{Z}_{\text{MC}}\hat{X}^{\otimes (n+m)}
= \hat{I} - 2\left(\ket{0}\bra{0}\right)^{\otimes (n+m)}$, the circuit implements
$\hat{D}_m$ up to the global phase $-1$ in Eq.~\eqref{Dmm}. For qudits of dimension $d>2$, the
Hadamard gate is replaced by the generalized Hadamard (quantum Fourier transform)
$\hat{H}_d$, with $\hat{H}_d^{\dagger}\neq\hat{H}_d$, and the $X$-conjugated
multi-controlled Z on the qudit registers is replaced by a phase gate controlled on each
qudit being in $\ket{0}$.

This whole sequence reflects all amplitudes around their average and, in turn, amplifies the amplitude of the marked state. Let $P_{\mathbf{y}_s}$ be the probability of measuring the subspace $\ket{\mathbf{y}}_s\ket{1}^{\otimes m}$ in the full state of the system $\ket{\psi_3}$. After one iteration of the Grover operator, the probability increases from $P_{\mathbf{y}_s}$ to $\sin^2{(3\cdot \arcsin{P_{\mathbf{y}_s}^{1/2})}}$, which for $P_{\mathbf{y}_s}<<1$ is $\approx 9 P_{\mathbf{y}_s}$. For the amplitude amplification to reach probability $1$ of measuring the marked entangled state $\ket{\mathbf{y}}_s\ket{1}^{\otimes m}$, one has to take into account both registers. Thus, the Grover operator $\hat{G}_m=\hat{D}_m\hat{O}_m$ is iteratively applied to the qubit register and keeps amplifying the target state. After $p$ iterations of $\hat{G}_m$, the probability of the target state increases to $\sin^2{((2p+1)\cdot \arcsin{P_{\mathbf{y}_s}^{1/2}})}$, which for small probabilities $P_{\mathbf{y}_s} << 1$, is $\approx \sin^2{((2p+1)\cdot P_{\mathbf{y}_s}^{1/2})}$.  The optimal number of Grover iterations needed to reach the probability $1$ of measuring the feasible region is $p_{\text{opt}} \approx (\pi/4)\cdot P_{\mathbf{y}_s}^{-1/2}$. In the worst-case scenario, when there is only one state in the feasible region, $P_{\mathbf{y}_s} = 1/d^{n}$, leading to the number of iterations $p_{\text{opt}} \approx (\pi/4)\cdot d^{n/2}$. The number of gates for this operation, counting two times cost of $\hat{U}_f$, two times cost of $\hat{H}_d$ and $\sim d^{n/2}/\sqrt{N_{\mathbf{y}_s}}$ as the number of Grover iterations, scales as $O(2\cdot m\cdot n^2\cdot n\cdot \log{d} \cdot d^{n/2}/\sqrt{N_{\mathbf{y}_s}} \cdot \log{\epsilon_G^{-1}})\sim O( d^{n/2}/\sqrt{N_{\mathbf{y}_s}})\text{ (in leading order terms) }$, which is exponential with the number of qudits $n$ and logarithmic in the precision $\epsilon_G$ \cite{brassard2000quantum,lin2022lecture}, where $N_{\mathbf{y}_s}$ is the number of multi-qudit states in the feasible region.

Finally, measuring the system will result in the state $\ket{1}^{\otimes m}$, and the system ends up in the following state,
\begin{equation}
\begin{split}
\begin{aligned}
  \ket{\psi_4} = F_{n}\sum_{\mathbf{y}_s\in\mathcal{S}} \ket{\mathbf{y}}_s\ket{1}^{\otimes m},
    \end{aligned}
\end{split}
\label{4}
\end{equation}
where $F_n = N_{\mathbf{y}_s}^{-1/2}$, with $N_{\mathbf{y}_s} \leq d^n$. $F_n$ is the normalizing factor, and the qudit states are collapsed to the feasible region of the problem.

\subsection{Stage II: Maximizing the cost function}
\label{stage2}

\noindent \textit{Step 4. Mid-circuit initialization.}
After following the steps of the constraint-satisfying stage that takes the multi-qudit states from a large Hilbert space to a relatively smaller subspace $\ket{\mathbf{y}}_s$ forming the feasible region, the subsequent stage is to maximize the cost function $C(\mathbf{y})$ in the feasible region. The constraint register, consisting of $m$ qubits, is obsolete after the measurement. For the optimization procedure, we introduce $l+1$ qubits; all in the state $\ket{0}$, divided into two registers, one with $l$ qubits (further discussion is provided in Step 5) and the other with $1$ qubit. The new state of the system is given as,
\begin{equation}
\begin{split}
\begin{aligned}
  \ket{\psi_5} = F_{n}\sum_{\mathbf{y}_s\in\mathcal{S}} \ket{\mathbf{y}}_s\ket{0}^{\otimes l}\ket{0}.
    \end{aligned}
\end{split}
\label{5}
\end{equation}
To find the maximum cost function, the $ l$-qubit register will store the normalized cost function values, entangled with the qudit register containing the corresponding integer variables, while the single qubit will be used to optimize the cost function.

The Hadamard gates $\hat{H}_2$ as shown in Fig.~\ref{Dalg}, are applied to the $ l$-qubit register to prepare an equal superposition state,
\begin{equation}
\begin{split}
\begin{aligned}
\ket{\psi_6} = \hat{H}_2^{\otimes l}\ket{\psi_5}
&= F_{n}\sum_{\mathbf{y}_s} \ket{\mathbf{y}}_s \Bigg(\hat{H}_2^{\otimes l}\ket{0}^{\otimes l}\Bigg)\ket{0}\\
&= \tilde{F}_{n} \sum_{\mathbf{y}_s} \ket{\mathbf{y}}_s\Bigg(\sum_{k=0}^{2^l-1}\ket{k}\Bigg) \ket{0},
    \end{aligned}
\end{split}
\label{6}
\end{equation}
where $\tilde{F}_n = F_{n}\cdot \sqrt{2^{-l}}$ is the modified normalization factor for the state. $l$ such Hadamard gates are required.

\noindent \textit{Step 5. Encoding the cost function using quantum phase estimation.}
The goal is to find the multi-qudit state corresponding to the maximum value of the cost function $C(\mathbf{y}_s)$, which is denoted as $\ket{\mathbf{y}_s^*}$.
The cost function value is encoded as the phase of the corresponding multi-qudit state; the phase operator is denoted by $\hat{O}_c$. The application of the operator $\hat{O}_c$ to the qudit register of $\ket{\psi_6}$ reads,
    \begin{equation}
\begin{split}
\begin{aligned}
  \ket{\tilde{\psi}_6} &=  \tilde{F}_{n} \Bigg( \sum_{\mathbf{y}_s\in\mathcal{S}}  \hat{O}_c \ket{\mathbf{y}}_s\Bigg) \Bigg(\sum_{k=0}^{2^l-1}\ket{k}\Bigg) \ket{0} \\
  &=
  \tilde{F}_{n} \Bigg(\sum_{\mathbf{y}_s\in\mathcal{S}}  e^{i2\pi \tilde{\phi}(\mathbf{y}_s)} \ket{\mathbf{y}}_s\Bigg) \Bigg(\sum_{k=0}^{2^l-1}\ket{k}\Bigg) \ket{0},
    \end{aligned}
\end{split}
\label{6_tilde}
\end{equation}
extracting the cost function value in the phase of each qudit state. Any linear or non-linear polynomial-cost function can be implemented by combining Eqs.~\eqref{4_1}, \eqref{C1}, and~\eqref{C2}, given in Appendix~\ref{Prereq}. Here, we choose a non-linear cost function for illustration.

Consider a cost function of the form $C(\mathbf{y}) = a_1x_1x_2 + a_2x_3^2+a_3x_2^2x_3x_4^2$, where $\mathbf{y} \equiv \{x_1,x_2,x_3,x_4\}$ and $\ket{\mathbf{y}} = \ket{x_1,x_2,x_3,x_4}$.
The phase operator $\hat{O}_c$ encoding the cost function using $\hat{q}\ket{\mathbf{x_j}} = x_j\ket{\mathbf{x_j}}$ from Eq.~\eqref{4_1} is given, as follows,
\begin{widetext}
\begin{equation}
\begin{split}
\begin{aligned}
  \hat{O}_c \ket{\mathbf{x_1}}&\otimes \ket{\mathbf{x_2}} \otimes \ket{\mathbf{x_3}}\otimes \ket{\mathbf{x_4}}  \\& = e^{i2\pi/C_{ub}} e^{i2\pi a_3 (\hat{I} \otimes \hat{q}^2 \otimes \hat{q} \otimes \hat{q}^2)/C_{ub}} e^{i2\pi a_2 (\hat{I} \otimes \hat{I} \otimes \hat{q}^2 \otimes \hat{I})/C_{ub}}e^{i2\pi a_1 (\hat{q} \otimes \hat{q} \otimes \hat{I} \otimes \hat{I})/C_{ub}} \\& \quad\quad\quad\quad\quad\quad\quad\quad\Bigg(\ket{\mathbf{x_1}}\otimes \ket{\mathbf{x_2}} \otimes \ket{\mathbf{x_3}}\otimes \ket{\mathbf{x_4}}\Bigg) \\
  &=
  e^{i2\pi/C_{ub}}e^{i2\pi a_3 x_2^2x_3x_4^2/C_{ub}}e^{i2\pi a_2 x_3^2/C_{ub}} e^{i2\pi a_1 x_1x_2/C_{ub}} \\& \quad\quad\quad\quad\quad\quad\quad\quad\Bigg(\ket{\mathbf{x_1}}\otimes \ket{\mathbf{x_2}} \otimes \ket{\mathbf{x_3}}\otimes \ket{\mathbf{x_4}}\Bigg) \\
  &= e^{i2\pi (1+a_1x_1x_2 + a_2x_3^2+a_3x_2^2x_3x_4^2)/C_{ub}} \Bigg(\ket{\mathbf{x_1}}\otimes \ket{\mathbf{x_2}} \otimes \ket{\mathbf{x_3}}\otimes \ket{\mathbf{x_4}}\Bigg) \\
  &= e^{i2\pi [1+C(\mathbf{y})]/C_{ub}}\Bigg(\ket{\mathbf{x_1}}\otimes \ket{\mathbf{x_2}} \otimes \ket{\mathbf{x_3}}\otimes \ket{\mathbf{x_4}}\Bigg).
    \end{aligned}
\end{split}
\label{nle}
\end{equation}
where the operator is applied only to the states in the qudit register and the quantity $[1+C(\mathbf{y}_s)]/C_{ub} < 1$. The factor $e^{i2\pi/C_{ub}}$ implements the constant shift in $\tilde{\phi}$ and is retained in the controlled applications of $\hat{O}_c$. $C_{ub}$ is the upper bound for the cost function required to make sure the phase lies in $[0,2\pi]$, which can be calculated in polynomial time using classical algorithms.

For the first term in the cost function $a_1x_1x_2$, the phase operator is defined using Eq.~\eqref{4_1} that is $\hat{q}\ket{\mathbf{x_j}} = x_j\ket{\mathbf{x_j}}$ and the properties given by Eqs.~\eqref{C1} and~\eqref{C2}. The operator is thus,
\begin{align}
    e^{i2\pi a_1 (\hat{q} \otimes \hat{q} \otimes \hat{I} \otimes \hat{I})/C_{ub}}\ket{\mathbf{x_1}}\otimes \ket{\mathbf{x_2}} \otimes \ket{\mathbf{x_3}}\otimes \ket{\mathbf{x_4}} = e^{i2\pi a_1 x_1x_2/C_{ub}}\ket{\mathbf{x_1}}\otimes \ket{\mathbf{x_2}} \otimes \ket{\mathbf{x_3}}\otimes \ket{\mathbf{x_4}}.
\end{align}

The second term of the cost function $a_2x_3^2$ requires an additional application of $\hat{q}$, leading to $\hat{q}\hat{q}\ket{\mathbf{x_j}} = x_j^2\ket{\mathbf{x_j}}$, the corresponding operation is as follows,
\begin{align}
e^{i2\pi a_2 (\hat{I} \otimes \hat{I} \otimes \hat{q}^2 \otimes \hat{I})/C_{ub}}\ket{\mathbf{x_1}}\otimes \ket{\mathbf{x_2}} \otimes \ket{\mathbf{x_3}}\otimes \ket{\mathbf{x_4}} = e^{i2\pi a_2 x_3^2/C_{ub}}\ket{\mathbf{x_1}}\otimes \ket{\mathbf{x_2}} \otimes \ket{\mathbf{x_3}}\otimes \ket{\mathbf{x_4}}.
\end{align}

For the implementation of the last term $a_3x_2^2x_3x_4^2$, all of the above properties are utilized to give,
\begin{align}
e^{i2\pi a_3 (\hat{I} \otimes \hat{q}^2 \otimes \hat{q} \otimes \hat{q}^2)/C_{ub}}\ket{\mathbf{x_1}}\otimes \ket{\mathbf{x_2}} \otimes \ket{\mathbf{x_3}}\otimes \ket{\mathbf{x_4}}= e^{i2\pi a_3 x_2^2x_3x_4^2/C_{ub}}\ket{\mathbf{x_1}}\otimes \ket{\mathbf{x_2}} \otimes \ket{\mathbf{x_3}}\otimes \ket{\mathbf{x_4}}.
\end{align}
\end{widetext}

An example for defining the operator $\hat{O}_c$ corresponding to a linear cost function is given in Appendix~\ref{Prereq} along with its application to the multi-qudit state of the system.

The phase operator is used to perform a quantum phase estimation (QPE) procedure using Hamiltonian simulation of $\hat{O}_c$, which operates on the superposition state $\ket{\psi_6}$. This stores the phase (the normalized cost function) corresponding to each multi-qudit state in the $ l$-qubit register (QPE register). Then, a controlled rotation of the single qubit converts the eigenvalues of the $l$-qubit register to the amplitude of the state and finds the optimal solution, which will be further detailed in Step 6.
Our approach is inspired by the Harrow--Hassidim--Lloyd (HHL) algorithm \cite{harrow2009quantum}. However, it differs from the HHL in terms of the definition of the phase operator $\hat{O}_c$, leading to the procedure of encoding non-linear terms in the phase.

The QPE operator using $\hat{O}_c$ is constructed as,
\[\sum_{k=0}^{2^l-1}\ket{k}\bra{k} \otimes \hat{O}_c^k \text{ with } \hat{O}_c^k \ket{\mathbf{y}}_s =(e^{i2\pi \tilde{\phi}(\mathbf{y}_s)})^k\ket{\mathbf{y}}_s.\]
Applying this operator to  Eq.~\eqref{6}, results in,
\begin{equation}
\begin{split}
\begin{aligned}
\ket{\psi_7} &= \Bigg(\sum_{k=0}^{2^l-1}\ket{k}\bra{k} \otimes \hat{O}_c^k\Bigg)\ket{\psi_6} \\
&=  \tilde{F}_{n} \sum_{\mathbf{y}_s} \sum_{k=0}^{2^l-1} e^{i2\pi \tilde{\phi}(\mathbf{y}_s) k} \ket{k}\ket{\mathbf{y}}_s\ket{0},
    \end{aligned}
\end{split}
\label{7}
\end{equation}
where $\tilde{\phi}(\mathbf{y}_s) = \phi(\mathbf{y}_s) + 1/C_{ub}$ and $\phi(\mathbf{y}_s) = C(\mathbf{y}_s)/C_{ub}$. $C_{ub} = C_{cont} + 1.5$ where $C_{cont}$ is calculated in polynomial time using classical algorithms by solving the relaxed problem (integer to continuous variables).
The maximum value of $C(\mathbf{y}_s)$ is $C_{cont}$, leading to $\phi(\mathbf{y}_s) \in  [0,1)$ and $\tilde{\phi}(\mathbf{y}_s) \in (0,1)$. By construction, the values of $\tilde{\phi}(\mathbf{y}_s)$ are strictly less than $1$ to make sure that no information is lost during the QPE routine. This is because the eigenvalues of any phase unitary operator lie on a unit circle and are inherently periodic; that is, for integer $p$, $e^{i2\pi\theta} = e^{i2\pi(\theta+p)}$. Hence, in QPE, the integer part of the phase is not stored in the qubit register.
The lower bound on $\tilde{\phi}(\mathbf{y}_s)$ is made strictly greater than zero by adding $1/C_{ub}$ to $\phi(\mathbf{y}_s)$; the reason will be reflected later during the controlled single qubit rotation step.
The complexity for the QPE procedure is $O(n\cdot\epsilon_{\text{QPE}}^{-1})$, where $\epsilon_{\text{QPE}}$ is the precision for the QPE operation \cite{brassard2000quantum,mande2023tight}.

Lastly, for the QPE routine, the inverse QFT operation represented as $\hat{U}^{\dagger}_{QFT}$, is performed to extract the value of the phase into the $l$-qubit register, given as
\begin{equation}
\begin{split}
\begin{aligned}
\hat{U}^{\dagger}_{QFT} \ket{k} &= \frac{1}{\sqrt{2^l}} \sum_{j=0}^{2^l - 1} e^{-2\pi i \frac{j k}{2^l}} \ket{j}
\\
\ket{\psi_8} &=  \hat{U}^{\dagger}_{QFT} \ket{\psi_7}=F_{n} \sum_{\mathbf{y}_s}  \ket{\tilde{\phi}(\mathbf{y}_s)}_l \ket{\mathbf{y}}_s\ket{0},
    \end{aligned}
\end{split}
\label{8}
\end{equation}
which requires $O(l^2)$ gate operations \cite{nielsen2010quantum}. The intermediate steps for the above inverse QFT operation are given in Appendix~\ref{psi8}.
$\ket{\tilde{\phi}(\mathbf{y}_s)}_l$ is the $l$-qubit superposition state which sharply peaks at the binary representation of the corresponding value of the phase $\tilde{\phi}(\mathbf{y}_s)$.
The operator $\hat{U}^{\dagger}_{QFT}$ is implemented using Hadamards and controlled rotation gates \cite{weinstein2001implementation}.

\noindent \textit{Step 6. Amplitude encoding of the cost function.} The qubit register in Eq.~\eqref{8} encodes the cost function values. To find the maximum one, the eigenvalues are encoded in the amplitude of the full state.
This is done by a controlled rotation $\hat{R}$ on the ancilla qubit based on the eigenvalue of the $l$-qubit register. The inverse of the eigenvalues $\tilde{\phi}(\mathbf{y}_s)$ ($\neq 0$) is encoded in the amplitude of the qubit state as $C/\tilde{\phi}(\mathbf{y}_s)$ for the $\ket{1}$ and $\sqrt{1-C^2/\tilde{\phi}(\mathbf{y}_s)^2}$ for $\ket{0}$, where $C$ is a normalization constant for the rotation, which is chosen to be $1/C_{ub}$. Under exact cost encoding, the resulting state of the system, with the data register written first, is given as,
\begin{widetext}
 \begin{equation}
\begin{split}
\begin{aligned}
\ket{\psi_9} &= F_{n} \sum_{\mathbf{y}_s}\ket{\mathbf{y}}_s\ket{\tilde{\phi}(\mathbf{y}_s)}_l\hat{R}(\tilde{\phi}(\mathbf{y}_s)) \ket{0}  \\
&= F_{n} \sum_{\mathbf{y}_s}\ket{\mathbf{y}}_s \ket{\tilde{\phi}(\mathbf{y}_s)}_l\Bigg(\sqrt{1-\frac{1}{(1+C(\mathbf{y}_s))^2}}\ket{0} + \frac{1}{(1+C(\mathbf{y}_s))} \ket{1}\Bigg).
    \end{aligned}
\end{split}
\label{9.1}
\end{equation}
\end{widetext}
The complexity for this rotation is $O(\log{\epsilon_R^{-1}})$ for $\epsilon_R$ precision.
Here, the amplitude corresponding to the qubit state $\ket{1}$ is inversely proportional to the value of the cost function $C(\mathbf{y}_s)$; hence, the amplitude of the state $\ket{0}$ increases with $C(\mathbf{y}_s)$. The integer variables that maximize the cost function are encoded in the qudit states, while the cost function value is encoded in the $l$-qubit state entangled with the $\ket{0}$ of the ancilla qubit.

\noindent \textit{Step 7. Final measurement for the optimal solution.}
The amplitude of the states in Eq.~\eqref{9.1} is dependent on the value of the cost function of the problem. The measurement of the ancilla qubit with the desired state being $\ket{0}$ (pre-factor) provides the optimal solution with relatively high probability. If the measurement leads to $\ket{1}$, then the procedure has to be repeated until the state $\ket{0}$ is measured. The state after the measurement is,
\begin{equation}
\begin{split}
\begin{aligned}
\ket{\psi_{10}} =  &\frac{F_{n}}{\sqrt{p_0}} \sum_{\mathbf{y}_s} \sqrt{1 - \frac{1}{(1 + C(\mathbf{y}_s))^2}} \cdot \ket{\mathbf{y}}_s \ket{\tilde{\phi}(\mathbf{y}_s)}_l
    \end{aligned}
\end{split}
\label{10}
\end{equation}
where $p_0 = |F_n|^2\sum_{\mathbf{y}_s} \left(1 - 1/(1 + C(\mathbf{y}_s))^2 \right)$.
The beauty of our construction is such that a finite set of measurements of the $l$-qubit register will correspond to the optimal cost function value, with relatively high probability due to the phase-amplitude encoding. This can be observed from the equation above.
Before Stage II, all the states in the collapsed qudit register that satisfy the constraints are uniformly distributed, each with amplitude \(N_{\mathbf{y}_s}^{-1/2}\), where \(N_{\mathbf{y}_s}\) is the number of such states.
However, the amplitude of the state corresponding to the minimum value of the cost function \(C(\mathbf{y}_s)=0\) becomes zero.
This amplitude is distributed with a bias on the maximum cost function value among the remaining states, which maximizes the probability of the optimal solution. The qudit register collapses to the state encoding the optimal variables with the highest probability compared to the other states after this measurement. The query complexity of the algorithm for achieving a success probability $P$ is discussed next along with the bounds on the number of repetitions required.

\section{Complexity analysis and the success probability of the algorithm}
\label{COMPLEXAN}

\noindent \textit{Complexity analysis.} Table~\ref{analysis} presents a detailed breakdown of the quantum resources and time complexity associated with each stage of our quantum algorithm for solving the IP problem \cite{berry2014exponential,lin2022lecture,brassard2000quantum,mande2023tight,weinstein2001implementation}. Each step corresponds to a subroutine(s), where resources are categorized by gate type (single-/multi-qubit and single-/multi-qudit). And the time complexity is analysed in terms of the number of integer variables \(n\), constraints \(m\), qudit dimension \(d\), QPE register size \(l\), and algorithmic precision parameters \(\epsilon_G\), \(\epsilon_{\text{QPE}}\), and \(\epsilon_R\).

The algorithm begins with state initialization in Step 1, where \(n\) single-qudit gates are in a uniform superposition of the input variables, which scales linearly with \(n\), with a time complexity of \(O(n)\). Step 2 encodes constraints through sparse unitary operators, requiring \(O(m)\) multi-qudit gates. There is a logarithmic dependency on the dimension, leading to the overall time complexity being \(O(m \cdot n^2 \cdot \log d)\).
Step 3 is a computationally expensive subroutine: amplitude amplification, requiring exponential resources. Specifically, it involves \(\sim 6 \cdot d^{n/2}\) single-qudit and \(\sim d^{n/2}\) multi-qudit gates, resulting in a time complexity of \(O(d^{n/2}/\sqrt{N_{\mathbf{y}_s}} \cdot \log \epsilon_G^{-1})\). This step represents the algorithm's bottleneck in terms of gate counts, time complexity scaling, and the number of measurements required.

In Step 4, a series of \(l\) single-qubit gates are applied, scaling linearly as \(O(l)\), for QPE preparation.
In Step 5, Quantum Phase Estimation (QPE) is applied, using multi-qudit and multi-qubit gates. The associated time complexity is \(O(n \cdot \epsilon_{\text{QPE}}^{-1})\), showing the sensitivity of the process to precision. This step also consists of the inverse QFT procedure, with \(O(l^2)\) time scaling.
Step 6 implements a rotation, requiring only a single multi-qubit gate and an associated logarithmic cost \(O(\log \epsilon_R^{-1})\).
Step 7 incurs no resource cost.

\begin{table*}[t!]
\centering
\resizebox{1\textwidth}{!}{%
\begin{tabular}{|m{0.4cm}|m{1.6cm}|m{3.8cm}|m{3.8cm}|m{7cm}|} 
\hline
		\multicolumn{2}{|c|}{Steps}&\multicolumn{2}{c|}{Resource complexity}&\multicolumn{1}{c|}{Time complexity $T_Q$}\\
			\hline
			& Sub-steps & Qubit gates (single + multi) & Qudit gates (single + multi)&   \\
			\hline
            1.&$\ket{\psi_1} \mapsto \ket{\psi_2}$&$0$&$n + 0$&$O(n)$ \\
            \hline
            2.&$\ket{\psi_2} \mapsto \ket{\psi_3}$&$0$&$0+O(mn^2)$&$O(m\cdot n^2\cdot \log{d})$\\
            \hline
            3.&$\ket{\psi_3} \mapsto \ket{\psi_4}$&$0$&$\sim6\cdot d^{n/2}+ d^{n/2}$&$O(2\cdot m\cdot n^2\cdot n\cdot \log{d} \cdot d^{n/2}/\sqrt{N_{\mathbf{y}_s}} \cdot \log{\epsilon_G^{-1}})$\\
            \hline
            \multirow{2}{*}{4.}&$\ket{\psi_4} \mapsto \ket{\psi_5}$&$0$&$0$&$0$\\

            &$\ket{\psi_5} \mapsto \ket{\psi_6}$&$l + 0$&$0$&$O(l)$\\
            \hline
            \multirow{2}{*}{5.}&$\ket{\psi_6} \mapsto \ket{\psi_7}$&$0+l$&$0+O(n)$&$O(n\cdot \epsilon_{\text{QPE}}^{-1})$\\
            &$\ket{\psi_7} \mapsto \ket{\psi_8}$&$l+l$&$0$&$O(l^2)$\\
            \hline
            6.&$\ket{\psi_8} \mapsto \ket{\psi_9}$&$0+1$&$0$&$O(\log{\epsilon_R^{-1}})$\\
            \hline
            7.&$\ket{\psi_9} \mapsto \ket{\psi_{10}}$&$0$&$0$&$0$\\
            \hline
            &\textbf{Total}&$(\sim 2l) + (2l+ 1)$&$n+O(n+mn^2) + 6 \cdot d^{n/2} + d^{n/2}$&$O(n+m\cdot n^2\cdot \log{d}+ 2\cdot m\cdot n^2\cdot n\cdot \log{d} \cdot d^{n/2}/\sqrt{N_{\mathbf{y}_s}} \cdot \log{\epsilon_G^{-1}}+l+ l^2+n\cdot \epsilon_{\text{QPE}}^{-1}+\log{\epsilon_R^{-1}})$, in leading order $\approx O(m\cdot n^2\cdot \log{d} +d^{n/2}/\sqrt{N_{\mathbf{y}_s}}+n\cdot \epsilon_{\text{QPE}}^{-1})$\\
            \hline
\end{tabular}%
}
        	\caption{\label{analysis}Step-wise resource and time complexity analysis of the quantum algorithm for a single run. Each transformation $\ket{\psi_i} \mapsto \ket{\psi_{i+1}}$ corresponds to a specific subroutine for which single- and multi-qubit/qudit gates, and the associated computational time are calculated. Resource complexity is given in terms of the gate types, while time complexity is expressed as a function of qudits ($n$), qubits ($m$, $l$), qudit dimension $d$, size of the feasible subspace $N_{\mathbf{y}_s}$, and precision parameters $\epsilon_G$, $\epsilon_{\text{QPE}}$, and $\epsilon_R$. The last row indicates the total resource counts and provides an asymptotic time complexity dominated by leading-order terms.}
\end{table*}

The total time complexity of the algorithm across all steps in a single run is
\begin{widetext}
\begin{equation}
    \begin{split}
        O\bigg(n+m\cdot n^2\cdot \log{d}+ 2\cdot m\cdot n^2\cdot n\cdot \log{d} \cdot d^{n/2}/\sqrt{N_{\mathbf{y}_s}} \cdot \log{\epsilon_G^{-1}}+l+ l^2+n\cdot \epsilon_{\text{QPE}}^{-1}+\log{\epsilon_R^{-1}}\bigg),
    \end{split}
\end{equation}
\end{widetext}
to reach a state having the optimal solution with the highest relative probability of measurement among all the feasible states. In leading-order terms, neglecting lower-order constants and considering a small QPE register (low $l$), the time simplifies to:
\begin{equation}
O\left(m\cdot n^2\cdot \log d + d^{n/2} + n \cdot \epsilon_{\text{QPE}}^{-1}\right).
\end{equation}

This expression highlights the key contributions to the overall complexity: the encoding of the constraints, the exponential cost of amplitude amplification, and the precision overhead of the QPE procedure. Since the algorithm depends on the size of the feasible region, let $N_{\mathbf{y}_s} \in [1,d^{n}]$ denote the size of the feasible region, leading to the time complexity of:
\begin{equation}
\boxed{
O\left(m\cdot n^2\cdot \log d + \frac{d^{n/2}}{\sqrt{N_{\mathbf{y}_s}}} + n \cdot \epsilon_{\text{QPE}}^{-1}\right).}
\end{equation}
The upper bound on $N_{\mathbf{y}_s}$ in our algorithm can be estimated from the unitary operators describing the distillation function, by counting the number of off-diagonal elements in each $\hat{U}_{f_i}$, giving,
\[
N_{\mathbf{y}_s} \leq \frac{1}{2} \min_i \#\left\{ (j, k) \mid j \ne k,\; [\hat{U}_{f_i}]_{jk} \ne 0 \right\},
\]
where the symbol $\#$ denotes `number of elements'. For this count, each $\hat{U}_{f_i}$ acts on the full $n$-qudit register and its single constraint flag, without spectator flags. Half its off-diagonal count equals the number of assignments satisfying constraint $i$ individually; the minimum of these counts is an upper bound on the number satisfying all constraints.

\noindent \textit{Success probability.} The process of resolving the optimal cost function is determined by the accuracy and the precision of the QPE procedure (thus the choice of quantum platform) that modifies the final probabilities before measurements \cite{brassard2000quantum,martinez2020quantum, van2021quantum}. Here, we provide bounds on the success probability depending on the accuracy of QPE and a relationship between the precision and the IP problem to gain more insights.
Let $\mathcal{S} \subset \mathbb{Z}^n$ denote the feasible region with $N_{\mathbf{y}_s} = |\mathcal{S}|$, and let $C(\mathbf{y}_s): \mathcal{S} \to \mathbb{R}_{\ge 0}$ be the cost function. $C_E(\mathbf{y}_s)$ is the QPE output of the cost function value $C(\mathbf{y}_s)$. Ideally, the quantum state at the end of the protocol should be given by Eq.~\eqref{9.1}. However, due to errors in QPE, the output state will contain estimates $\tilde{\phi}_E(\mathbf{y}_s)$ of $\tilde{\phi}(\mathbf{y}_s) = (C(\mathbf{y}_s) + 1)/C_{ub}$. For the bounded finite-QPE rotation convention given in Appendix~\ref{psi8}, use $C_E(\mathbf{y}_s)=\max\{0,C_{ub}\tilde{\phi}_E(\mathbf{y}_s)-1\}$. On the event that the phase error is at most $\delta_{\text{QPE}}$, the output and the ideal case give,
\begin{widetext}
\begin{equation}
    \begin{split}
        |\tilde{\phi}_E(\mathbf{y}_s) - \tilde{\phi}(\mathbf{y}_s)| &\le  \delta_{\text{QPE}} \\
      \max\{0,C(\mathbf{y}_s)-C_{ub}\cdot\delta_{\text{QPE}}\}\le C_E(\mathbf{y}_s)  &\le C(\mathbf{y}_s) +C_{ub}\cdot\delta_{\text{QPE}}.
    \end{split}
\end{equation}
\end{widetext}
This provides bounds to the output cost function value $C_E(\mathbf{y}_s)$, which is the same as the ideal cost function $C(\mathbf{y}_s)$ if $\delta_{\text{QPE}} = 0$.

Let \( \mathbf{y}_s^* = \arg\max_{\mathbf{y}_s\in \mathcal{S}} C(\mathbf{y}_s) \), with the cost function $C(\mathbf{y}_s^*)$, being the optimal solution.
Considering the amplitudes are uniform over the feasible region, i.e., \( |F_n|^2 = 1/N_{\mathbf{y}_s}\).
The joint success probability of the selected ancilla outcome and one optimal assignment $p$ for the solution in the ideal case ($\delta_{\text{QPE}} = 0$) is,
\begin{equation}
    p = \frac{1}{N_{\mathbf{y}_s}} \cdot \left(1 - \frac{1}{(1 + C(\mathbf{y}_s^*))^2} \right).
    \label{suP}
\end{equation}
For non-zero $\delta_{\text{QPE}}$, define $C_-^*=\max\{0,C(\mathbf{y}_s^*)-C_{ub}\delta_{\text{QPE}}\}$ and $C_+^*=C(\mathbf{y}_s^*)+C_{ub}\delta_{\text{QPE}}$. The joint success probability is bounded as:
\begin{equation}
\begin{split}
\frac{1}{N_{\mathbf{y}_s}}
\left(1-\frac{1}{(1+C_-^*)^2}\right)
\leq p\leq
\frac{1}{N_{\mathbf{y}_s}}
\left(1-\frac{1}{(1+C_+^*)^2}\right).
\end{split}
\label{prob2}
\end{equation}
The interval between the probability bounds increases as $\delta_{\text{QPE}}$ is increased, leading to further uncertainty in the final solution.

Let $\mathbf{y}_s^{\dagger} \in \mathcal{S}$ be a near-optimal solution with cost function $C(\mathbf{y}_s^{\dagger})$. Then, to resolve the probability between the optimal $\mathbf{y}_s^*$ and the sub-optimal solution $\mathbf{y}_s^{\dagger}$, the difference between the QPE estimates has to be more than the precision $\epsilon_{\text{QPE}}$, leading to the condition,
\begin{equation}
    \begin{split}
        \frac{C_E(\mathbf{y}_s^*) - C_E(\mathbf{y}_s^{\dagger})}{C_{ub}} > \epsilon_{\text{QPE}},
    \end{split}
\end{equation}
when taking the lower limit of $C_E(\mathbf{y}_s^*)$ and upper limit of $C_E(\mathbf{y}_s^{\dagger})$ gives the following condition,
\begin{equation}
    \begin{split}
        C(\mathbf{y}_s^*) - C(\mathbf{y}_s^{\dagger}) > \epsilon_{\text{QPE}}\cdot C_{ub}  + 2C_{ub}\cdot\delta_{\text{QPE}}.
    \end{split}
    \label{prec}
\end{equation}
On the stated QPE error events, this condition is sufficient to resolve the estimated costs corresponding to the optimal and the near-optimal solutions. If the precision of QPE $\epsilon_{\text{QPE}}$ attains larger values, this condition will not be satisfied at a certain threshold, and hence, the probabilities will show degeneracies. In general, $\epsilon_{\text{QPE}}$ is given by the number of noiseless qubits $l$ in the QPE as $1/2^l$, which depends on the experimental platform.

\noindent \textit{Number of Repetitions.}
The quantum algorithm outputs the optimal solution with probability $p$, given by Eq.~\eqref{suP}.
To increase the probability of finding the optimal solution, the algorithm needs to be repeated \(r\) times independently \cite{tanaka2022noisy,brassard2000quantum}. The probability of failure (no success) in all \(r\) runs is
\[
(1 - p)^r.
\]
Hence, the probability of obtaining at least one success in \(r\) repetitions is
\[
P(\text{at least one success}) = 1 - (1 - p)^r.
\]

\begin{figure}[tbp]
\centering
\includegraphics[width = 1\linewidth,trim={4.7cm 10.2cm 4.7cm 10.3cm},clip]{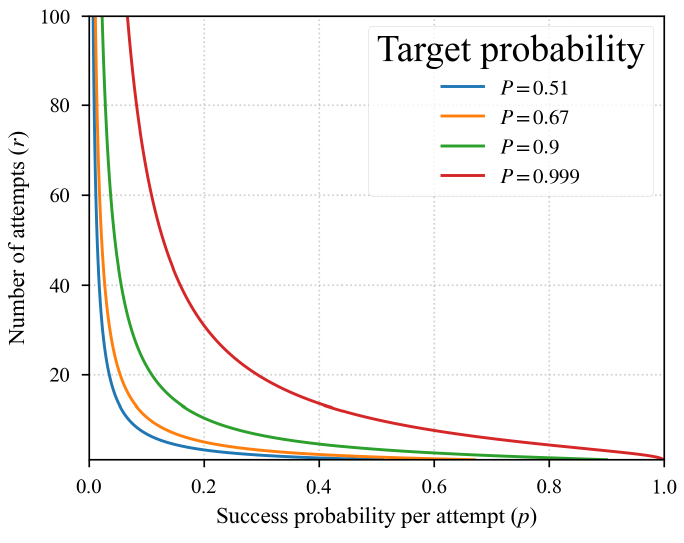}
\caption{
The relationship between the number of attempts ($r$) and the required success probability per attempt ($p$) to achieve various target probabilities ($P$).
The curves show that as the success probability per attempt decreases, the required number of attempts increases.
For a given target probability $P$, the success probability $p$ is $p = 1 - (1 - P)^{1/r}$.
The plot shows that more attempts allow for lower individual success probabilities while still achieving the same overall target probability.
}
\label{r_vs_p}
\end{figure}

In order to guarantee this probability to be at least a target threshold \(P\), then
\begin{equation}
    \begin{split}
        &\quad \quad 1 - (1-p)^r \geq P \\
        & \implies
       (1-p)^r \leq 1 - P, \\
      & \implies r \cdot \log(1-p) \leq \log(1 - P).
    \end{split}
\end{equation}

Since \(\log(1-p) < 0\), rearranging yields
\begin{equation}
r \geq r_{\min}=\left\lceil\frac{\log(1-P)}{\log(1-p)}\right\rceil,\qquad 0<p<1,\quad 0<P<1.
\label{repitions}
\end{equation}

Fig.~\ref{r_vs_p} shows the number of repetitions required for a given success probability to reach a target success probability of $0.51$, $0.67$, $0.80$, and $0.999$. All of the target probabilities are such that the algorithm provides the optimal solution more than $50\%$ of the time, and shows that with $50-100$ repetitions, many of the problems (with $p \leq 0.1$) can be solved with high probability.

For small \(p\) and when \(P\) is close to $1$, using the approximation \(\log(1-x) \approx -x\) for small \(x\),
\begin{equation}
    r \approx \frac{-\log(1 - P)}{p}.
\end{equation}

For example, for \(P=0.99\),
$-\log(1 - 0.99) = -\log(0.01) \approx 4.605.$
And so the number of repetitions required and using Eq.~\eqref{suP} is
\begin{equation}
\boxed{
r \approx \frac{4.605}{p} = \frac{4.605 \cdot N_{\mathbf{y}_s}}{\left(1 - \frac{1}{(1 + C(\mathbf{y}_s^*))^2} \right)}.}
\end{equation}
The number of repetitions to get to the optimal solution depends on the problem, which can be estimated using classical pre-processing of the problem. Using the branch and bound algorithm \cite{beale1965mixed} can provide a lower bound on the cost function value, which can bound the number of repetitions required. As the classical algorithm reports the first feasible solution, using that as the lower bound for $C(\mathbf{y}_s^*)$ leads to the maximum repetition $r$ for a particular problem that guarantees the solution using the quantum algorithm. Hence, the time complexity for solving an IP problem using our algorithm with a success probability $P$ requiring $r_P$ repetitions is,
\begin{equation}
\boxed{
O\left(r_P\cdot \big(m\cdot n^2\cdot \log d + \frac{d^{n/2}}{\sqrt{N_{\mathbf{y}_s}}} + n \cdot \epsilon_{\text{QPE}}^{-1}\big)\right).}
\end{equation}

\begin{figure*}[tp]
    \centering
    \includegraphics[width = 1\linewidth,trim={10cm 0cm 12cm 0cm},clip]{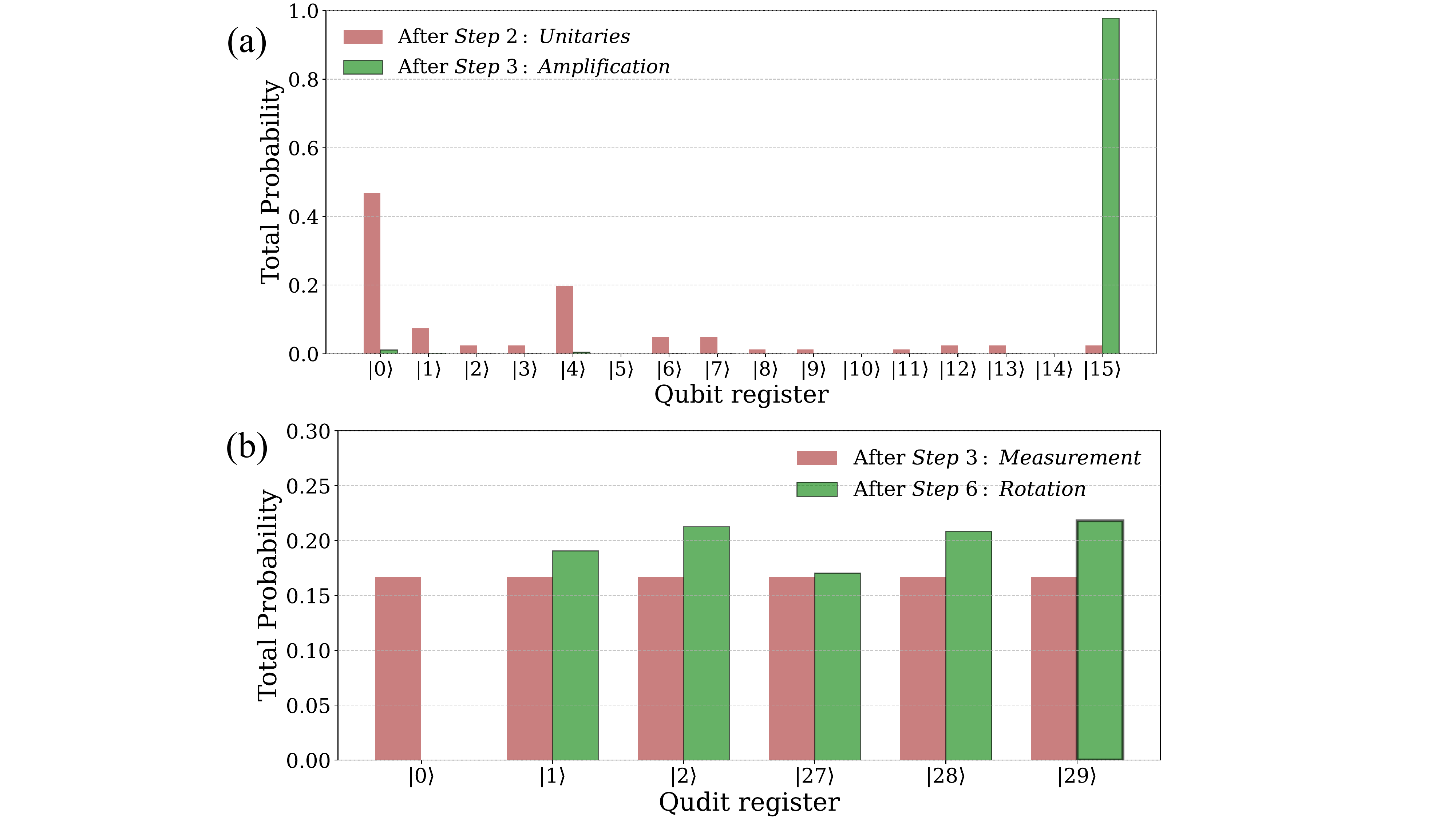}
    \caption{Quantum solution of a non-linear integer program with 5 variables and 4 non-convex constraints. In both panels, the states on the x-axis are numbered basis states, for (a) it ranges from $0$ to $2^4-1$, and for (b) only the states in the feasible region are shown.
(a) Shows qubit register (constraint) probabilities before (brown) and after (green) amplitude amplification, highlighting the amplification of the feasible state $\ket{15}$.
(b) Shows qudit register probabilities before (brown) and after (green) cost function optimization via QPE, with the optimal solution $\ket{\mathbf{29}}$ (cost $= 4$) amplified.
In the ideal cost-encoding model, $19$ accepted samples with ancilla outcome $0$ give a success probability of at least $0.99$. }
    \label{NLIP}
\end{figure*}

\section{\label{results} Numerical demonstration of the qudit-based quantum algorithm}
The quantum algorithm presented in this work can tackle any linear and non-linear IP problem; however, non-linear problems are much harder to solve classically \cite{jeroslow1973there}. This is because often non-linear terms in constraints make the optimization non-convex \cite{wolsey1999integer}. Hence, for a demonstration that our quantum algorithm can tackle classically hard problems, we chose a sample non-linear IP given as,
\begin{equation}
\begin{split}
C(\mathbf{x}) &=2 x_1 + x_2 +  x_3 + 3x_4 + \frac{3}{2}x_5 \\
\text{To } &\text{be maximized under:} \\
C_1&: x_1+x_2^2x_3+x_3 <1  \\
C_2&:3x_3^2x_4+x_2 < 2 \\
C_3&: x_1x_5+x_4 <1  \\
C_4&:2x_1+2x_1^2x_3 + x_4^3 < 2 \\
x_i &\in \{0,1,2\}.
\end{split}
\label{ExP1}
\end{equation}
and solve it using our algorithm. The problem contains quadratic and cubic terms, making it difficult for general-purpose gradient-based and interior-point methods to solve \cite{wright2005interior}, precisely due to the non-convexity of the problem. Often, the brute-force method is preferred for solving such problems. The non-convexity of the problem is analyzed in Appendix~\ref{convex}.
The algorithm is implemented using Python \cite{python}, where each operator shown in Fig.~\ref{Dalg} is well-defined mathematically, leading to the execution of the corresponding circuit architecture for solving the above IP problem.

Fig.~\ref{NLIP} shows the result of the implementation of the quantum algorithm for solving this non-linear non-convex optimization problem. The probability distribution of the constraint register in panel (a) and the data register in panel (b) is shown during the algorithmic run. The algorithm initialises with $5$-qudits corresponding to $5$ variables, each taking $3$-values and $4$-qubits for the constraint register. The total number of qudit states is $3^5 = 243$. In panel (a), the probability of the basis of many-qubit states (numbered from $0$ to $2^4 - 1$) is shown, where each basis has two bars corresponding to the probability after applying entangling unitary operators $U_{f_i}$ (brown bars) and after the amplitude amplification step (green bars) respectively. The unitary operators (as described in Step 2 in Sec.~\ref {Alg}) give bias to some of the basis states, $\ket{0}$ in this case, indicating that many of the qudit states in the superposition are entangled with $\ket{0}$, a consequence of a large infeasible region. To amplify the qubit state $\ket{1111}$ (represented as $\ket{15}$ in panel (a) of Fig.~\ref{NLIP}) entangled with the feasible region (all constraints satisfied), a Grover operator is applied (Step 3 in Sec.~\ref{Alg}), maximizing the probability of the target state, as shown by the green bar.

After measuring the qubit register, the corresponding qudit register collapses to the states satisfying all the constraints simultaneously, which amounts to a total of $6$ out of $3^5 (= 243)$ states as shown in panel (b). The brown bars in panel (b) show an equal superposition of the states in the feasible region, with the basis being numbered. The goal now is to find the one state that maximizes the cost function. As described in Sec.~\ref {Alg}, a QPE procedure with $l=4$ qubits is performed, and another ancilla qubit is rotated based on the QPE register. This results in the final redistributed probabilities in the feasible region, as shown by green bars in panel (b). There are three key messages here: (i) the probability of the optimal cost function is always higher than the rest, indicated by the highlighted green bar with black cover, (ii) in the ideal cost-encoding model, $p_0\approx0.734311$, the joint probability of the optimum and ancilla outcome $0$ is $p_{*,0}=(1-1/25)/6=0.16$, and the conditional probability is $p_{*\mid0}=p_{*,0}/p_0\approx0.217891$. Equation~\eqref{repitions} then gives $19$ accepted samples with ancilla outcome $0$, or $27$ attempts at the cost-filtering stage starting from the uniform feasible state, to reach $P_{target}=0.99$; the finite-$l$ calculation uses the same formula with its actual probability, (iii) In the ideal cost-encoding model, the probability of the qudit state $\ket{\mathbf{0}}$ after the cost function optimization procedure goes to zero, and this excess probability is distributed among the other states with a bias based on the cost function value.
In this case, the state $\ket{\mathbf{29}} \equiv \ket{01002}$ is the optimal solution with $C(x) = 4$.

\begin{figure*}[tp]
    \centering
  \includegraphics[width = 1\linewidth,trim={5.2cm 0cm 4.3cm 0cm},clip]{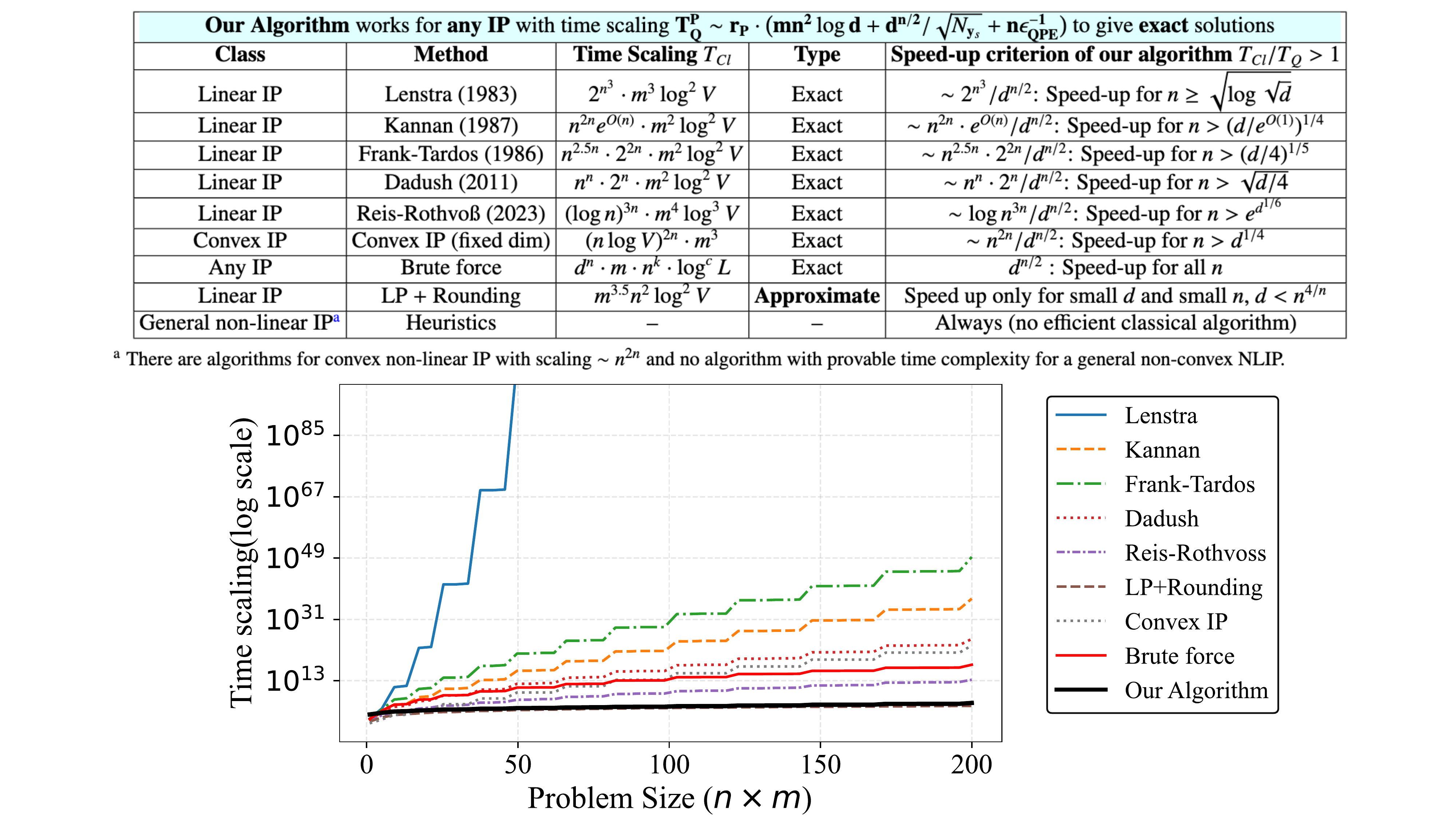}
    \caption{Comparative analysis of integer programming algorithms. Top: Time complexity expressions for each method, along with the class of problems that can be solved. The table also shows the criterion when our algorithm has a speed-up over the classical cases for solving IP. For most of the algorithms, as $n$ increases, the quantum algorithm becomes faster, except for the approximate LP+Rounding, which is used to solve linear problems. The approximation ratio for LP+rounding algorithms is bounded by the integrality gap (ratio of the optimal solution of the relaxed-continuous variable problem and the IP problem), which is $2$ for the vertex cover problem, $\ln{n}$ for the set cover problem, and $3/4$ for the Max-SAT problem \cite{thai2013approximation}. All logarithms are base-2. V is the maximum coefficient magnitude, $k$ is the degree of the polynomials, $c$ is the bit-precision, and $\epsilon_{\text{QPE}}$ is the quantum phase estimation precision.  Bottom: Time complexity scaling with problem size ($n \times m$) with $d=3$, $r_P = 10$, $N_{\mathbf{y}_s} = 10$ $V=10$, $c= 2$, $k=3$ and $\epsilon_{\text{QPE}} = 0.1$.}
    \label{Complex}
\end{figure*}

For comparison, a branch and bound method is used to solve the same problem, which provides the optimal solution; however, it requires $358$ nodes (computations). The brute-force method requires $3^5 (= 243)$ computations, which is less than that of the branch and bound algorithm. We also solved the problem using a brute force approach and Google-OR-tools CP-SAT solver \cite{perron2024or}; the brute force required one-tenth of the time to reach the solution compared to the CP-SAT solver. This indicates that as the non-convexity in the problem increases, brute force becomes better than the available general-purpose solvers. Hence, for such integer programming problems, it suffices to compare the time complexity with the brute-force method. However, the mentioned classical methods (other than the brute-force) do work well for convex problems. We compare next the time complexity of various special-purpose classical algorithms with our algorithm, where a discussion of the limitations of classical algorithms is also presented.

In Fig.~\ref{Complex}, the time complexity scaling of multiple classical algorithms, along with the brute force approach, is shown and compared with our quantum algorithm. The classical exact algorithms presented in the top table show a provable time scaling with various parameters of the IP problem; however, the methods are limited to linear IPs \cite{lenstra1983integer,kannan1987minkowski,frank1987application,dadush2012,reisrothvoss2023,eisenbrand2008parametric,tawarmalani2013convexification}.
This is due to these algorithms relying on the optimization landscape being a convex polytope, which allows them to use geometric methods, such as using ellipsoids to find the solution. On the other hand, approximate algorithms for solving IP show polynomial time scaling with problem size, but with multiple assumptions, such as in fully polynomial-time approximation schemes (FPTAS) \cite{fptas-knapsack}, only a single linear constraint with a quadratic cost function is allowed, or in linear-programming relaxation with rounding (LP+rounding), it fails for non-linear constraints and suffers from rounding off errors.
There are multiple commercially available solvers for IP, such as Gurobi, CPLEX, BARON, Couenne, and IOPT \cite{kilincc2018exploiting,andersen2000mosek,meindl2012analysis}, with underlying algorithms being either one or a combination of branch and bound, cutting planes, and heuristics, coupled with the relaxation of the integer constraint. To solve IP using these algorithms, a relaxation step is used to convert the integer variables to continuous, and then the relaxed problem is solved with modifications in each iteration. Most of these commercially available classical solvers also have a few restrictions on the class of problems they can solve efficiently, such as the constraints must be linear and the inequalities must be non-strict.
The non-strict inequalities (such as $C_i \leq h_i$ instead of $C_i < h_i$) are to make sure that there are no open sets in the optimization polytope. The solution of the relaxed problems resides on the boundaries of this polytope, and open sets remove the best guess for the relaxed problems.
The table in Fig.~\ref{Complex} also indicates the threshold where our algorithm shows a speed-up over the classical algorithms listed for solving IP. The brute force algorithm starts performing better than the other classical algorithms for higher $n$ as well; however, it always scales worse than the quantum algorithm ($d^n$ vs $(\sqrt{d})^n$), and its time complexity increases further with the degree of the polynomial $k$ (non-linearity of the IP), leading to a larger advantage in our approach. The last row of the table indicates that, in practice, heuristics are often used to solve a given non-convex IP problem, but they fail to guarantee the optimal solution or provide an analytical time scaling.

In the bottom panel in Fig.~\ref{Complex}, we compare the time complexity of the various algorithms listed in the top table with our algorithm as a function of problem size $n\times m$, as both $n$ and $m$ contribute to the complexity of the problem. For the plot, the dimension of the variables is $d=3$, the largest coefficient in the problem $V$ is chosen to be $10$, and $\epsilon_{\text{QPE}}$ is $0.1$, which is typical in digital quantum circuits. $V$ is the maximum coefficient that appears either in the constraints or the cost function, for example, in Eq.~\eqref{ExP1}, $V=3$; the scaling of many algorithms depends on this parameter due to all the numbers being stored in bits. For computational purposes, a threshold of $10^{100}$ for the y-axis is fixed, as it is expensive to show further. At $n\times m =50$, the Lenstra algorithm reaches this threshold. The scaling of our quantum algorithm with problem size is by far the best compared to the other exact algorithms, and it beats the approximate algorithms in terms of the class of problems our algorithm can tackle. The algorithm addresses bounded nonlinear polynomial-IP problems, which are NP-hard in general \cite{hemmecke2009nonlinear}. Without finite bounds on the integer variables, polynomial-IP can be undecidable, even with quadratic constraints \cite{jeroslow1973there}. The bounded formulation considered here is decidable by exhaustive enumeration of the $d^n$ candidate assignments. One of the features of our algorithm is that there is an intermediate measurement step after the constraint-satisfying stage, which helps to mitigate such infinite loops practically. If in that step, the probability of measuring $\ket{1}^{\otimes m}$ is zero, the amplitude amplification will not result in the state $\ket{1}^{\otimes m}$, leading to the conclusion that the feasible region for the given constraints does not exist.
In that case, the computation can be stopped after a few measurements, as the problem might be infeasible. For practical purposes, this can save computational resources and identify the difficulty of solving the given problem instance. The output of the measurement of the constraint register also contains information about the states that satisfy a range of constraints, which can help in identifying the maximum number of constraints required to make the problem feasible. For example, if probability of $\ket{1}^{\otimes m}$ is zero and the next non-zero probability multi-qubit state has one of the qubits is in $\ket{0}$ while the rest are in $\ket{1}$, then the problem may become feasible if one of the constraints is removed. Counting the number of zeros in the subsequent multi-qubit states provides the crucial information for making the problem feasible. Next, we analyse our algorithm further to understand the limits for achieving a quantum advantage.

\begin{figure}[tbp]
    \centering
  \includegraphics[width = 1\linewidth,trim={5.5cm 0cm 4cm 0cm},clip]{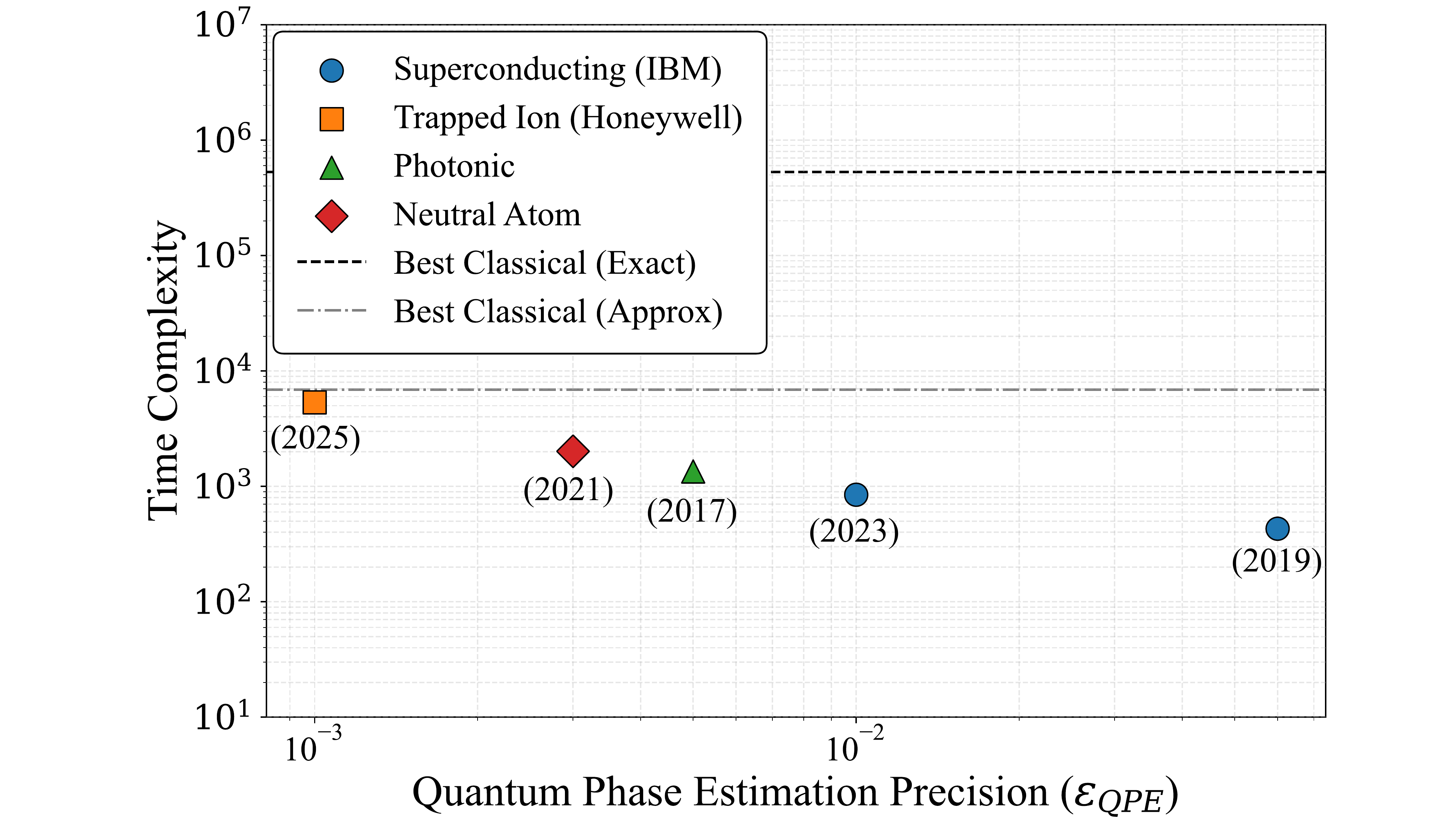}
    \caption{
Comparison of time complexity versus quantum phase estimation precision \(\epsilon_{\mathrm{QPE}}\) for our quantum algorithm across various experimental platforms: Superconducting (2019,2023) \cite{mohammadbagherpoor2019improved,blunt2023statistical}, Trapped Ion (2025) \cite{yamamoto2025quantum}, Photonic (2017) \cite{paesani2017experimental}, and Neutral Atom (theory with experimental parameters, 2021) \cite{pezze2021quantum}. Each data point corresponds to reported \(\epsilon_{\mathrm{QPE}}\) values and demonstration years. The time complexity for a single run, computed as \(m n^2 \log_2 d + d^{n/2}/\sqrt{N_{\mathbf{y}_s}} + \frac{n}{\epsilon_{\mathrm{QPE}}}\) with \(m = n = d = 5, N_{\mathbf{y}_s} = 5\), is plotted on a log-log scale. Horizontal dashed and dash-dot lines represent the complexities of one of the best-known classical exact algorithms (Reis--Rothvoss) and one of the best approximate classical algorithms (LP + rounding), respectively.
}
    \label{QPE}
\end{figure}

Fig.~\ref{QPE} analyses the time complexity of our quantum algorithm with the precision $\epsilon_{\text{QPE}}$ reached by various quantum computing platforms. The plot considers superconducting (experiment), trapped ion (experiment), photonic (experiment), and neutral atoms (theory with experimental considerations) quantum computing platforms \cite{pezze2021quantum,paesani2017experimental,yamamoto2025quantum,mohammadbagherpoor2019improved,blunt2023statistical}, where $\epsilon_{\text{QPE}}$ is characterized by the errors in the gates and the number of qubits in the register used for the experiment. So far, there have been a few experiments for QPE, and for the platforms where the precision is not explicitly provided, it is estimated by the number of qubits used in the procedure. For $l$-qubits in QPE, the smallest phase that can be determined is given by $1/2^l$, which defines the precision for the procedure. In general, as the number of error-corrected qubits increases, the precision gets better.

For the plot, the following parameters were chosen: $n=m=d=5$, and for classical algorithms, $V=10$. The two horizontal lines show the time complexity of the best exact classical algorithm and the best approximate classical algorithm, indicating that if a problem with such parameters is considered, the quantum algorithm can get competitive for polynomial repetitions. It is also crucial to the quantum algorithm that an exponentially precise phase estimation is not a strict requirement, as the separation of the relative phases between the multi-qudit states is enough to maximize the optimal cost function. The only condition to be satisfied is given by Eq.~\eqref{prec}, which ensures that the quantum algorithm provides a valid solution. The smaller the RHS of Eq.~\eqref{prec}, the better it is for the quantum algorithm. However, a smaller $\epsilon_{\text{QPE}}$ leads to higher time complexity; for all practical purposes, $\epsilon_{\text{QPE}}\sim 10^{-3}$ \cite{lim2024curve} is a good compromise for the given parameters. If the problem size increases, the classical algorithms scale worse as shown in Fig.~\ref{Complex}, allowing for a smaller $\epsilon_{\text{QPE}}$ to retain the quantum advantage. If a non-linear problem such as the one given by Eq.~\eqref{ExP1} is solved, then the classical algorithms may run into difficulties. However, heuristics that perform global optimization for nonlinear and commercial solvers can still be comparable to the quantum algorithm, but for a limited class of problems.

\section{\label{conc}Discussion and conclusions}
Integer Programming (IP) is an optimization framework with widespread applications in logistics, finance, manufacturing, and decision-making under constraints \cite{wolsey1999integer}. Despite the sophisticated classical algorithms, IP remains computationally intractable for large problem sizes due to its NP-hard nature \cite{papadimitriou1982complexity}. This motivates the development of quantum algorithms that can provide a quantum advantage over classical methods, either through reduced complexity, improved scalability, or both.
Most existing quantum approaches for combinatorial optimization problems, such as IP, rely on qubits or variational methods, which are often expensive in terms of resources. The lack of efficient encodings for integer variables has been a bottleneck in applying quantum algorithms to IP.

Building on prior work that solved small-scale IP using a single qudit \cite{goswami2024integer}, this work extends the algorithm to arbitrary IP problems by introducing a complete quantum circuit that achieves both constraint satisfaction and cost-function optimization. The measurement at the end yields the maximum probability for the optimal solution(s) by construction, leading to fewer circuit runs to achieve a high success probability as compared to a randomized algorithm.
Our approach offers a speed-up in time complexity for solving IP over the exhaustive search, from \(T_c \sim O(d^n)\) for the classical case to
\begin{equation}
\boxed{
T^P_Q = O\left(r_P\cdot \big(m\cdot n^2\cdot \log d + \frac{d^{n/2}}{\sqrt{N_{\mathbf{y}_s}}} + n \cdot \epsilon_{\text{QPE}}^{-1}\big)\right).}
\end{equation}
where \(n\) is the number of variables, each taking \(d\) values, \(m\) the number of constraints, $r_P$ repetitions to achieve a success probability of $P$ and \(\epsilon_{\text{QPE}}\) the QPE precision. This is the best-known analytical scaling for a general bounded polynomial IP, which constitutes the non-convex non-linear cases as well, and sets a new benchmark for quantum optimization algorithms.
The subroutines in our algorithm are all performed using only quantum operations, which include the optimization of the classical cost function, hence shielding itself from the issues faced by variational quantum algorithms \cite{bittel2021training,bittel2022optimizing}.
Due to the measurement-based approach in the quantum algorithm, for practical purposes, this can distinguish between infeasible and feasible problem instances before performing the cost function optimization. The qubit register stores the information about the constraint satisfiability and the number of satisfied constraints, which can help to modify the problem by removing one or more constraints to make the problem feasible. All operations are implemented within a gate-based framework, providing a path toward fault tolerance, thereby motivating the experiments \cite{fedorov2022quantum}.

For the practical realization of our algorithm, the essential components required are qudits, qudit/qubit gates, and hybrid qubit-qudit entangling gates. The qudits can be realised in a quantum system with a large number of controllable degrees of freedom, for example hyperfine/electronic states in atoms \cite{kanungo2022realizing,gadway2015atom} or non-trivial superposition of them like Rydberg-dressed states \cite{macri2014rydberg,mukherjee2019charge}, frequency modes in photonic system \cite{ozawa2019topological,yuan2018synthetic}, rotational states in ultracold polar molecules \cite{shaffer2018ultracold,sawant2020ultracold,gadway2016strongly} and the simulation of \textit{synthetic} dimensions \cite{boada2012quantum,martin2017topological,sundar2018synthetic}. Furthermore, the field for experimental demonstration of qudit gates is also an active area, to list a few, a qutrit ($d=3$) CZ gate on superconducting qubits \cite{Goss2022}, a native 2-qudit entangling gate up to $d=5$ dimensions on trapped ions \cite{Hrmo2023}, a 2-ququarts ($d=4$) cross-resonance entangling gate using transmon-based architecture \cite{Fischer2023}, and a qudit C-NOT gate for $d=4$ on NV center platform \cite{Du2024}. Formulation for universal qudit gate sets (Pauli, Hadamard, Z-rotation, controlled, and SWAP gates) for implementing QFT \cite{Pudda2024} is also presented, along with a proposal for hybrid gates acting on qudits of different dimensions (2-qudit SUM, SWAP, many-qudit hybrid Toffoli and Fredkin gates) on spin systems
\cite{Daboul2003}. However, for proof of principle of our algorithm, the near-term devices can solve a binary non-linear IP with the current gate sets available, which, for a highly non-convex problem, can potentially indicate an advantage over the classical algorithms.

As with all current quantum algorithms, there can be various sources of error. One of the significant issues that can arise is due to the errors in the unitary operator representing the distillation function that implements the constraints. This can cause states in the qudit register to leak out of the feasible region, leading to states from the infeasible region interfering with the optimization procedure. Despite this, the optimal solution remains measurable with a greater number of iterations. Once the integer values are decoded, the satisfaction of the constraints can be checked in polynomial time, and the next probable integer assignment can be chosen as the solution. The other sources of errors in the circuit, such as during the amplitude amplification and the QPE, are well studied with statements about their robustness \cite{li2021errorQPE,meier2019testing,azuma2002decoherence,salas2008noise,pomeransky2004groverImperfections,zhirov2006dissipative,shapira2003unitaryNoise}.
Further study on the effect of experimental noise in our algorithm is an outlook for us.
Also, hybrid algorithms combining quantum operations with classical post-processing could further improve the efficiency of the algorithm. Other IP problems containing trigonometric functions, logarithms, and rational polynomials ($P(X)/Q(X)$) in the cost function require further studies to be tackled within our framework.

\begin{acknowledgments}
This work is funded by the German Federal Ministry of Education and Research within the funding program ``Quantum Technologies -- from basic research to market'' under Contract No.~13N16138.
\end{acknowledgments}

\appendix

\section{Generalization of the Hadamard gates for \texorpdfstring{$d$}{d}-dimensional qudits}
\label{Hadamards}
The Hadamard gate for a qudit with $d$ dimensions is defined as
 \begin{equation}
\begin{split}
\begin{aligned}
\hat{H}_d\ket{\mathbf{\alpha}} = \frac{1}{\sqrt{d}}\sum_{\beta=0}^{d-1} (e^{2\pi\text{i}/d})^{\alpha\beta}\ket{\mathbf{\beta}}, \alpha\in\{0,1,...,d-1\},
    \end{aligned}
\end{split}
\label{Hp}
\end{equation}
for a state $\ket{\mathbf{\alpha}}= \ket{\mathbf{0}}$, it gives,
\[\hat{H}_d\ket{\mathbf{0}} = \frac{1}{\sqrt{d}}\sum_{\beta=0}^{d-1}\ket{\mathbf{\beta}}.\]
$n$ such Hadamard gates are used for the step $\ket{\psi_1} \rightarrow \ket{\psi_2}$ as given by Eq.~\eqref{2} in the main text.

\section{Devising the distillation function and its operator implementation}
\label{fdis}

To define the distillation function $f_i$ for a constraint $C_i(y^i)<h_i$ acting on a
subset of the integer variables $y^i$, $f_i$ has to follow the condition that it separates
the states forming the feasible and the infeasible region. The required operation to be
performed on a computational basis state is
\begin{widetext}
\begin{equation}
\begin{split}
\begin{aligned}
  \ket{\mathbf{y},0}^i \rightarrow \hat{U}_{f_i} \ket{\mathbf{y},0}^i
  &= \ket{\mathbf{y},0 \oplus_2 f_i(y^i)}^i
  = \begin{cases}
      \ket{\mathbf{y},0}_k^i & : \text{Infeasible region} \\
      \ket{\mathbf{y},1}_s^i & : \text{Feasible region},
    \end{cases}
\end{aligned}
\end{split}
\label{equ}
\end{equation}
where the operator $\hat{U}_{f_i}$ performs a Boolean addition $\oplus_2$, i.e., addition
modulo 2, of the distillation function $f_i$ to the qubit register, such that
$0\oplus_2 f_i = f_i \bmod 2$. Hence,
\begin{equation}
\hat{U}_{f_i}\sum_{\mathbf{y}} \alpha_{\mathbf{y}}\ket{\mathbf{y},0}^i
= \sum_{\mathbf{y}\,\in\,\text{infeasible}} \alpha_{\mathbf{y}}\ket{\mathbf{y},0}_k^i
+ \sum_{\mathbf{y}\,\in\,\text{feasible}} \alpha_{\mathbf{y}}\ket{\mathbf{y},1}_s^i .
\label{equ_sup}
\end{equation}
\end{widetext}
The function $f_i$ depends on the constraint $C_i<h_i$ and is given by Eq.~\eqref{fun} in
the main text, defined as
\begin{equation}
f_i(y^i) = 2^{\floor{\frac{C_i(y^i)}{h_i}}},
\label{fun2}
\end{equation}
where we assume $C_i\geq 0$ and $h_i>0$. This can always be achieved by shifting the
constraint by the minimum value $C_i^{\min}$ of $C_i$, i.e., $C_i\rightarrow C_i-C_i^{\min}$
and $h_i\rightarrow h_i-C_i^{\min}$; if $h_i\leq C_i^{\min}$, no assignment satisfies the
constraint. The function is such that, if the constraint is satisfied, the value of $f_i$
is 1; otherwise, it is an even integer, as shown below.
\begin{equation}
\begin{split}
\begin{aligned}
    \text{If } & \begin{cases}
      \frac{C_i}{h_i} \geq 1 & : \text{Infeasible region}  \\
      0 \leq \frac{C_i}{h_i} < 1 & : \text{Feasible region}
    \end{cases} \\
    \text{then } & \begin{cases}
      \floor{\frac{C_i}{h_i}} \in \mathbb{Z}^+ & : \text{Infeasible region}  \\
      \floor{\frac{C_i}{h_i}} = 0 & : \text{Feasible region}
    \end{cases} \\
   \implies & f_i \bmod 2 =  \begin{cases}
      0 \quad \left(2^{\floor{\frac{C_i}{h_i}}} \text{ even}\right) & : \text{Infeasible region}  \\
      1 \quad \left(2^{\floor{\frac{C_i}{h_i}}} = 1\right) & : \text{Feasible region}.
    \end{cases}
\end{aligned}
\end{split}
\label{equ2}
\end{equation}

From Eq.~\eqref{equ} along with the definition of $f_i$ in Eq.~\eqref{equ2}, the unitary
$\hat{U}_{f_i}$ can be constructed, which is given by Eq.~\eqref{equ31}.
For illustration of the matrix and the circuit for a $\hat{U}_{f_i}$, consider a simple
constraint $C_1: x_1 + 2x_2<2$, where $x_i\in\{0,1\}$. Its feasible region is
$\{(x_1,x_2)\}=\{(0,0),(1,0)\}$, i.e., the constraint is satisfied if and only if $x_2=0$.
In the basis $\ket{x_1 x_2\, q_0}$, the corresponding $\hat{U}_{f_1}$ is an $8 \times 8$
matrix that specifically maps the following states,
\begin{center}
$\ket{\mathbf{00}0} \gr{\xleftrightarrow{\text{Qubit flip}}} \ket{\mathbf{00}1}$\\
$\ket{\mathbf{01}0} \re{\xleftrightarrow{\text{Qubit unchanged}}} \ket{\mathbf{01}0}$\\
$\ket{\mathbf{10}0}\gr{\xleftrightarrow{\text{Qubit flip}}}\ket{\mathbf{10}1}$\\
$\ket{\mathbf{11}0}\re{\xleftrightarrow{\text{Qubit unchanged}}} \ket{\mathbf{11}0}$,\\
\end{center}
and the $\hat{U}_{f_1}$ is,
\begin{equation}
\hat{U}_{f_1} =
\begin{bmatrix}
0 & \gr{1} & 0 & 0 & 0 & 0 & 0 & 0 \\
\gr{1} & 0 & 0 & 0 & 0 & 0 & 0 & 0 \\
0 & 0 & \re{1} & 0 & 0 & 0 & 0 & 0 \\
0 & 0 & 0 & \re{1} & 0 & 0 & 0 & 0 \\
0 & 0 & 0 & 0 & 0 & \gr{1} & 0 & 0 \\
0 & 0 & 0 & 0 & \gr{1} & 0 & 0 & 0 \\
0 & 0 & 0 & 0 & 0 & 0 & \re{1} & 0 \\
0 & 0 & 0 & 0 & 0 & 0 & 0 & \re{1}
\end{bmatrix}
\label{U_Mat}
\end{equation}
which is a 1-sparse matrix. The circuit implementation using a C-NOT gate,
where the qubits $\mathbf{q_1}$ and $\mathbf{q_2}$ encode $x_1$ and $x_2$ and $q_0$ is the
flag qubit, is
\begin{center}
\begin{quantikz}
\lstick{$\mathbf{q_1}$} & \qw      & \qw      & \qw      & \qw \\
\lstick{$\mathbf{q_2}$} & \gate{X} & \ctrl{1} & \gate{X} & \qw \\
\lstick{$q_0$}          & \qw      & \targ{}  & \qw      & \qw
\end{quantikz}
\end{center}
which flips $q_0$ only when $q_2=0$, i.e., when $x_2=0$, requiring 3 gates (or a single
C-NOT with a control on $\ket{0}$).

Alternatively, $\hat{U}_{f_1}$ can be obtained from the $8\times8$ identity matrix by
permuting its rows according to the transpositions in Eq.~\eqref{U_Mat}, i.e., exchanging
rows $1\leftrightarrow2$ and $5\leftrightarrow6$. Each transposition exchanges two basis
states that differ only in the flag qubit $q_0$.

\section{Sequential application of \texorpdfstring{$m$}{m}-unitary operators}
\label{Sequential}

The resulting state after sequential operations of $m$-$\hat{U}_{f_i}$ is,
\begin{widetext}
\begin{equation}
\begin{split}
\begin{aligned}
\ket{\psi_3} &= \prod_{i=1}^{m}\hat{U}_{f_i}\ket{\psi_2}
=\frac{1}{d^{n/2}}\prod_{i=1}^{m}\hat{U}_{f_i}
\Bigg(\sum_{\mathbf{y}\in\{0,\ldots,d-1\}^n}\ket{\mathbf{y}}\ket{0}^{\otimes m}\Bigg)\\
&=\frac{1}{d^{n/2}}\sum_{\mathbf{y}\in\{0,\ldots,d-1\}^n}
\ket{\mathbf{y}}\ket{q(\mathbf{y})}\\
&=\frac{1}{d^{n/2}}\Bigg(
\sum_{\gamma=0}^{m-1}\sum_{\mathbf{y}\in\mathcal{K}_\gamma}
\ket{\mathbf{y}}\ket{q(\mathbf{y})}
+\sum_{\mathbf{y}_s\in\mathcal{S}}\ket{\mathbf{y}}_s\ket{1}^{\otimes m}\Bigg).
\end{aligned}
\end{split}
\label{3}
\end{equation}
\end{widetext}
Here, $\ket{q(\mathbf{y})}$ records which constraints are satisfied by each assignment, and has Hamming weight $\gamma$ for $\mathbf{y}\in\mathcal{K}_{\gamma}$. The states $\ket{\mathbf{y}}_s$ satisfy all $m$ constraints and are paired with $\ket{1}^{\otimes m}$.

\section{Grover amplitude amplification}
\label{Grovers}

The Grover operator is defined as $\hat{G} = \hat{D} \hat{O}$, where $\hat{O} = \hat{I} - 2 \ket{M} \bra{M}$ is the phase flip operator and $\hat{D} = 2\ket{\psi}\bra{\psi} - \hat{I}$ is the diffusion operator that performs a reflection around the mean. The goal is to maximize the amplitude of a given state $\ket{M}$ out of a superposition state $\ket{\psi}$.
Let's take an $m$-qubit case, where the state with all the qubits is state $\ket{1}$ is the target state. The operator $\hat{O}$ in this case is as: $\hat{O} = \hat{I} - 2 \ket{1}^{\otimes m} \bra{1}^{\otimes m}$, where \( \ket{1}^{\otimes m} \) is the target state we are searching for. This operation applies a phase flip to \( \ket{1}^{\otimes m} \) while leaving other states unchanged:
\[
\hat{O} \ket{1}^{\otimes m} = - \ket{1}^{\otimes m}, \quad \hat{O} |x\rangle = |x\rangle \text{ for } x \neq 11...1.
\]

The diffusion operator \( \hat{D} \) enhances the amplitude of the target state by reflecting all amplitudes around their mean. It is given by: $\hat{D} = 2 |\psi\rangle \langle \psi| - \hat{I}$,
where \( |\psi\rangle \) is the superposition state from which the target state needs to be sampled.
\begin{equation}
    |\psi\rangle = \frac{1}{\sqrt{N}} \sum_{x} |x\rangle.
\end{equation}

Here, $(\ket{1}^{\otimes m})^{\perp}$ is the normalized uniform superposition of the $N-1$ unmarked states, so that $\ket{\psi}=\sin\theta\ket{1}^{\otimes m}+\cos\theta(\ket{1}^{\otimes m})^{\perp}$.
Applying \( \hat{D} \) to the target state and its perpendicular state results in,
\begin{equation}
    \begin{split}
        \hat{D} \ket{1}^{\otimes m} &= -\cos(2\theta) \ket{1}^{\otimes m} + \sin(2\theta) (\ket{1}^{\otimes m})^{\perp}
\\
\hat{D} (\ket{1}^{\otimes m})^{\perp} &= \sin(2\theta) \ket{1}^{\otimes m} + \cos(2\theta) (\ket{1}^{\otimes m})^{\perp},
    \end{split}
\end{equation}
where \( \theta \) is given by $ \sin^2(\theta) = \frac{1}{N}$ and $N = 2^m$ in the case where $\ket{\psi}$ is in the uniform superposition of all the possible qubit states.
Since the Grover operator \( \hat{G} \) is a rotation in a 2D subspace spanned by \( \ket{1}^{\otimes m} \) and its orthogonal complement \( (\ket{1}^{\otimes m})^{\perp} \), its eigenstates are:
\begin{equation}
 |\psi_{\pm}\rangle = \frac{1}{\sqrt{2}} \left( \ket{1}^{\otimes m} \pm i (\ket{1}^{\otimes m})^{\perp} \right).
\end{equation}

The corresponding eigenvalues of \( \hat{G} \) are  $\lambda_{\pm} = e^{\pm 2i \theta}$.
Since the search space is large (\( N = 2^m \)), for small \( P_{11...1} = \frac{1}{N} \), we approximate: $\theta \approx \sqrt{P_{11...1}} = \frac{1}{\sqrt{N}}$.
Each application of \( \hat{G} \) rotates the initial uniform state \( |\psi\rangle \) toward \( \ket{1}^{\otimes m} \) by an angle \( 2\theta \). The number of iterations required to reach the first maximum of \( P_{11...1} \) is:

\begin{equation}
    p_{\text{opt}} \approx \frac{\pi}{4\theta}-\frac{1}{2} \sim \frac{\pi}{4}\sqrt{N}.
\end{equation}

The integer iteration count is chosen by rounding $\pi/(4\theta)-1/2$ to the nearest integer. For large $N$, the probability of measuring the target state at this first maximum approaches 1.

Now, we detail the stepwise application of the operator $\hat{O}_m$ to the state $\ket{\psi_3}$ in our algorithm, as
\begin{widetext}
\begin{align}
\hat{O}_m\ket{\psi_3}\ket{-}
 &= \frac{1}{d^{n/2}}\hat{O}_m\Bigg (\sum_{(k_\gamma,\gamma)} \ket{\mathbf{y}}_{k_\gamma}\ket{q(\mathbf{y}_{k_\gamma})} + \sum_{\mathbf{y}_s\in\mathcal{S}} \ket{\mathbf{y}}_s\ket{ 1}^{\otimes m}\Bigg)\ket{-} \nonumber\\
&= \frac{1}{d^{n/2}}\Bigg (\sum_{(k_\gamma,\gamma)} \ket{\mathbf{y}}_{k_\gamma}\hat{X}_{MC}\Big(\ket{q(\mathbf{y}_{k_\gamma})}\ket{-}\Big)  + \sum_{\mathbf{y}_s\in\mathcal{S}} \ket{\mathbf{y}}_s\hat{X}_{MC}\Big(\ket{ 1}^{\otimes m}\ket{-}\Big)\Bigg) \nonumber\\
&= \frac{1}{d^{n/2}}\Bigg (\sum_{(k_\gamma,\gamma)} \ket{\mathbf{y}}_{k_\gamma}\Big(\ket{q(\mathbf{y}_{k_\gamma})}\ket{-}\Big)  + \sum_{\mathbf{y}_s\in\mathcal{S}} \ket{\mathbf{y}}_s(-1)\Big(\ket{ 1}^{\otimes m}\ket{-}\Big)\Bigg)\nonumber\\
&= \frac{1}{d^{n/2}}\Bigg (\sum_{(k_\gamma,\gamma)} \ket{\mathbf{y}}_{k_\gamma}\ket{q(\mathbf{y}_{k_\gamma})} - \sum_{\mathbf{y}_s\in\mathcal{S}} \ket{\mathbf{y}}_s\ket{ 1}^{\otimes m}\Bigg)\ket{-}.
\label{O_m_apply2}
\end{align}
If the ancilla qubit is in state $\ket{0}$ instead of $\ket{-}$, applying a multi-controlled $Z$ ($\hat{Z}_{MC}$) with control over the qubit register will result in the same marked state. The $\hat{Z}_{MC}$ gate can be implemented as, $\hat{X}_{MC}(a_1, ..., a_m; a) \quad \hat{Z}(a) \quad \hat{X}_{MC}(a_1, ..., a_m; a),$
where $a$ represents the ancilla qubit and $(a_1 a_2 \cdots a_m)$ is the qubit register.

Let
\begin{equation}
    d^{-n/2}\sum_{(k_\gamma,\gamma)} \ket{\mathbf{y}}_{k_\gamma}\ket{q(\mathbf{y}_{k_\gamma})}\equiv\alpha_k\ket{\mathbf{K},Q}
\end{equation}
\begin{equation}
    d^{-n/2}\sum_{\mathbf{y}_s\in\mathcal{S}} \ket{\mathbf{y}}_s\ket{ 1}^{\otimes m}\equiv \beta_s\ket{\mathbf{S},1^{\otimes m}},
\end{equation}
then $\ket{\psi_3} =$ $\alpha_k\ket{\mathbf{K},Q}$ $+$ $\beta_s\ket{\mathbf{S},1^{\otimes m}}$.
And the diffusion operator $\hat{D}_m$ reflecting the state around the average is defined as,
\begin{align}
\hat{D}_m &= 2\ket{\psi_3}\bra{\psi_3} - \hat{I}, \nonumber \\
&= 2\bigg(\alpha_k\ket{\mathbf{K},Q} +\beta_s\ket{\mathbf{S},1^{\otimes m}}\bigg)\bigg(\alpha^*_k\bra{\mathbf{K},Q} +\beta^*_s\bra{\mathbf{S},1^{\otimes m}}\bigg) - \hat{I},
\end{align}
\end{widetext}
where $\ket{\psi_3}$ is the initial superposition state of the system before the application of the operator $\hat{O}_m$, $\alpha^*_k$ and $\beta^*_s$ are the complex conjugates of $\alpha_k$ and $\beta_s$, respectively.

\section{Encoding of the cost function as the phase of the state}
\label{Prereq}

A prerequisite for QPE is to encode the cost function $C(\mathbf{y}_s)$ for each multi-qudit state in the feasible region to the $l$-qubit register. To do this, the operator $\hat{q}$ is,
    \begin{equation}
    \hat{q}\ket{\mathbf{x_j}} = x_j\ket{\mathbf{x_j}},
    \label{4_1}
    \end{equation}
where $x_j$ is the eigenvalue of a single qudit state $\ket{\mathbf{x_j}}$. This eigenvalue operator is used to construct a unitary operator $\hat{O}_c$ for imprinting the cost as the phase of the corresponding many-body qudit state.
For any cost function of IP, we use the properties listed below for the implementation of $\hat{O}_c$.

Consider two sub-spaces of a system $A$ and $B$ with arbitrary operators, $\hat{H}_A$, $\hat{H}_B$ and Identity operators $\hat{I}_A$, $\hat{I}_B$.
Then, $e^{\hat{H}_A \otimes \hat{I}_B} = e^{\hat{H}_A} \otimes \hat{I}_B$, the proof is given below.
\begin{widetext}
     \begin{align}
  Proof&:e^{\hat{H}_A \otimes \hat{I}_B}
  =  \sum_{i=0}^\infty \frac{\Bigg(\hat{H}_A \otimes \hat{I}_B\Bigg)^i}{i!} \nonumber\\
  &= \Bigg(\hat{I}_A \otimes \hat{I}_B\Bigg) + \Bigg(\hat{H}_A \otimes \hat{I}_B\Bigg) + \frac{1}{2}\Bigg(\hat{H}_A \otimes \hat{I}_B\Bigg)^2 + ... \nonumber\\
  &= \Bigg(\hat{I}_A \otimes \hat{I}_B\Bigg) + \Bigg(\hat{H}_A \otimes \hat{I}_B\Bigg) + \frac{1}{2}\Bigg(\hat{H}_A^2 \otimes \hat{I}_B^2\Bigg) + ... \nonumber\\
  &= \Bigg(\hat{I}_A + \hat{H}_A  + \frac{1}{2}\hat{H}_A^2 + ...\Bigg)\otimes \hat{I}_B \nonumber\\
  &= e^{\hat{H}_A}\otimes \hat{I}_B.
\label{proof1}
\end{align}
For a system with multiple sub-spaces $A$, $B$, $C$ and $D$, it follows,
    \begin{equation}
    \begin{split}
    \begin{aligned}
    e^{\hat{I}_A \otimes \hat{H}_B \otimes \hat{I}_C \otimes \hat{I}_D} = \hat{I}_A \otimes e^{\hat{H}_B} \otimes \hat{I}_C \otimes \hat{I}_D.
           \end{aligned}
    \end{split}
    \label{C1}
    \end{equation}

Consider the following eigenvalue equations,
\[\hat{H}_A \ket{\psi_A} = \epsilon_A \ket{\psi_A} \text{ and } \hat{H}_B \ket{\psi_B} = \epsilon_B \ket{\psi_B} \]
 \begin{equation}
   \text{then: } e^{H_A \otimes H_B} \ket{\psi_A}\otimes \ket{\psi_B} = e^{\epsilon_A \epsilon_B} \ket{\psi_A}\otimes \ket{\psi_B}.
\end{equation}
And the proof is given below.
    \begin{align}
  Pro&of:e^{\hat{H}_A \otimes \hat{H}_B} \ket{\psi_A}\otimes \ket{\psi_B} \nonumber\\
&=  \sum_{i=0}^\infty \frac{\Bigg(\hat{H}_A \otimes \hat{H}_B\Bigg)^i}{i!} \ket{\psi_A}\otimes \ket{\psi_B} \nonumber\\
  &= \Bigg(\Bigg(\hat{I}_A \otimes \hat{I}_B\Bigg) + \Bigg(\hat{H}_A \otimes \hat{H}_B\Bigg) + \frac{1}{2}\Bigg(\hat{H}_A \otimes \hat{H}_B\Bigg)^2 + ...\Bigg)\ket{\psi_A}\otimes \ket{\psi_B} \nonumber\\
  &= \Bigg(\hat{I}_A \otimes \hat{I}_B\Bigg) \ket{\psi_A}\otimes \ket{\psi_B} + \Bigg(\hat{H}_A \otimes \hat{H}_B\Bigg) \ket{\psi_A}\otimes \ket{\psi_B}  + \frac{1}{2}\Bigg(\hat{H}_A^2 \otimes \hat{H}_B^2\Bigg) \ket{\psi_A}\otimes \ket{\psi_B} + ... \nonumber\\
  &= \Bigg(1 + \Bigg(\epsilon_A \epsilon_B\Bigg)  + \frac{1}{2}\Bigg(\epsilon_A \epsilon_B\Bigg)^2 + ...\Bigg)  \ket{\psi_A}\otimes \ket{\psi_B}\nonumber \\
  &= e^{\epsilon_A \epsilon_B} \ket{\psi_A}\otimes \ket{\psi_B}.
\label{proof2}
\end{align}

For a system with multiple sub-spaces,
\begin{equation}
\begin{split}
\begin{aligned}
e^{\hat{H}_A \otimes \hat{H}_B \otimes \hat{H}_C \otimes \hat{I}_D} \ket{\psi_A}\otimes \ket{\psi_B}\otimes \ket{\psi_C}\otimes \ket{\psi_D} \\= e^{\epsilon_A\epsilon_B\epsilon_C}\ket{\psi_A}\otimes \ket{\psi_B}\otimes \ket{\psi_C}\otimes \ket{\psi_D}.
       \end{aligned}
\end{split}
\label{C2}
\end{equation}
The encoding of the non-linear cost function using the above equations is given in the main text.

\noindent \textit{Step-wise encoding of a linear cost function as the phase of the state:}
Take a linear cost function of the form $C(\mathbf{y}_s) = a_1x_1 + \cdots + a_nx_n$. The eigenvalue operator given by Eq.~\eqref{4_1} is used to construct a unitary operator $\hat{O}_c$ for imprinting the cost as the phase of the corresponding many-body qudit state,
\begin{equation}
\hat{O}_c = e^{i2\pi/C_{ub}}\Big(e^{i2\pi\hat{q}a_1/C_{ub}} \otimes \cdots \otimes e^{i2\pi\hat{q}a_n/C_{ub}}\Big).
\label{4_22}
\end{equation}
The action of the operator $\hat{O}_c$ for the exemplar linear cost function for the state $\ket{\psi_6}$ is,
    \begin{align}
  \ket{\tilde{\psi}_6} &= \hat{O}_c \ket{\psi_6} \nonumber\\
  &=  \hat{O}_c  \Bigg(\tilde{F}_{n} \sum_{k=0}^{2^l-1}\ket{k}\Bigg) \Bigg(\sum_{\mathbf{y}_s\in\mathcal{S}} \ket{\mathbf{y}}_s\Bigg)\ket{0} \nonumber\\
  &= \tilde{F}_{n} e^{i2\pi/C_{ub}}\Bigg(e^{i2\pi\hat{q}a_1/C_{ub}} \otimes \cdots \otimes e^{i2\pi\hat{q}a_n/C_{ub}}\Bigg) \nonumber\\ &\quad \quad \quad \quad \Bigg(\sum_{\mathbf{y}_s\in\mathcal{S}} \ket{\mathbf{x_1}}_s \otimes \cdots \otimes \ket{\mathbf{x_n}}_s\Bigg) \otimes \Bigg( \sum_{k=0}^{2^l-1}\ket{k}\Bigg)\ket{0} \nonumber\\
  &= \tilde{F}_{n} e^{i2\pi/C_{ub}}\Bigg(\sum_{\mathbf{y}_s\in\mathcal{S}} e^{i2\pi\hat{q}a_1/C_{ub}} \ket{\mathbf{x_1}}_s \otimes \cdots \nonumber\\ & \quad \quad \quad \quad \quad \quad \otimes e^{i2\pi\hat{q}a_n/C_{ub}} \ket{\mathbf{x_n}}_s\Bigg) \otimes \Bigg(\sum_{k=0}^{2^l-1}\ket{k}\Bigg)\ket{0} \nonumber\\
  &= \tilde{F}_{n} \Bigg(\sum_{\mathbf{y}_s\in\mathcal{S}}  e^{i2\pi(1+a_1x_1^s + ... + a_nx_n^s )/C_{ub}} \nonumber\\ & \quad \quad \quad \quad \quad \quad  \Bigg(\ket{\mathbf{x_1}}_s \otimes \cdots  \otimes \ket{\mathbf{x_n}}_s\Bigg)\Bigg)\otimes \Bigg(\sum_{k=0}^{2^l-1}\ket{k}\Bigg)\ket{0} \nonumber\\
  &=
  \tilde{F}_{n} \Bigg(\sum_{\mathbf{y}_s\in\mathcal{S}}  e^{i2\pi [1+C(\mathbf{y}_s)]/C_{ub}} \ket{\mathbf{y}}_s\Bigg)\otimes \Bigg(\sum_{k=0}^{2^l-1}\ket{k}\Bigg)\ket{0},
\label{5_tilde}
\end{align}
\end{widetext}
where $\ket{\mathbf{x_i}}_s$ are the individual integer states of the qudit $i$ corresponding to a particular superposition state represented by $\ket{\mathbf{y}}_s$.

\section{Steps for the inverse QFT operation}
\label{psi8}

The state after imprinting the cost function as the phase is,

\begin{equation}
\ket{\psi_7}
=
\tilde{F}_n
\sum_{\mathbf{y}_s}
\sum_{k=0}^{2^l-1}
e^{i2\pi \tilde{\phi}(\mathbf{y}_s) k}
\ket{k}\ket{\mathbf{y}}_s\ket{0}.
\end{equation}

Apply the inverse QFT operator
$\hat{U}_{QFT}^{\dagger}$ to the $\ket{k}$ register,

\begin{equation}
\ket{\psi_8}
=
(\hat{U}_{QFT}^{\dagger} \otimes I)\ket{\psi_7},
\end{equation}

\begin{equation}
\ket{\psi_8}
=
\tilde{F}_n
\sum_{\mathbf{y}_s}
\sum_{k=0}^{2^l-1}
e^{i2\pi \tilde{\phi}(\mathbf{y}_s) k}
(\hat{U}_{QFT}^{\dagger}\ket{k})
\ket{\mathbf{y}}_s\ket{0}.
\end{equation}

Using,

\begin{equation}
\hat{U}_{QFT}^{\dagger}\ket{k}
=
\frac{1}{\sqrt{2^l}}
\sum_{j=0}^{2^l-1}
e^{-2\pi i \frac{jk}{2^l}}
\ket{j},
\end{equation}

we obtain

\begin{widetext}
\begin{equation}
\ket{\psi_8}
=
\tilde{F}_n
\sum_{\mathbf{y}_s}
\sum_{k=0}^{2^l-1}
e^{i2\pi \tilde{\phi}(\mathbf{y}_s) k}
\left(
\frac{1}{\sqrt{2^l}}
\sum_{j=0}^{2^l-1}
e^{-2\pi i \frac{jk}{2^l}}
\ket{j}
\right)
\ket{\mathbf{y}}_s\ket{0}.
\end{equation}
\end{widetext}

Rearranging the sums,

\begin{equation}
\ket{\psi_8}
=
\frac{\tilde{F}_n}{\sqrt{2^l}}
\sum_{\mathbf{y}_s}
\sum_{j=0}^{2^l-1}
\left(
\sum_{k=0}^{2^l-1}
e^{i2\pi k(\tilde{\phi}(\mathbf{y}_s)-j/2^l)}
\right)
\ket{j}\ket{\mathbf{y}}_s\ket{0}.
\end{equation}

The sum inside the big brackets is a geometric series,

\begin{equation}
\sum_{k=0}^{2^l-1}
e^{i2\pi k(\tilde{\phi}(\mathbf{y}_s)-j/2^l)} = \frac{1 - e^{i2\pi 2^l(\tilde{\phi}(\mathbf{y}_s)-j/2^l)}}
{1 - e^{i2\pi(\tilde{\phi}(\mathbf{y}_s)-j/2^l)}}.
\end{equation}

Hence $\ket{\psi_8}$,

\begin{equation}
\ket{\psi_8}
=
\frac{\tilde{F}_n}{\sqrt{2^l}}
\sum_{\mathbf{y}_s}
\sum_{j=0}^{2^l-1}
\frac{1-e^{i2\pi 2^l(\tilde{\phi}(\mathbf{y}_s)-j/2^l)}}
{1-e^{i2\pi(\tilde{\phi}(\mathbf{y}_s)-j/2^l)}}
\ket{j}\ket{\mathbf{y}}_s\ket{0}.
\end{equation}

The geometric series sharply peaks around
\begin{equation}
j \approx 2^l \tilde{\phi}(\mathbf{y}_s).
\end{equation}

In general, the phase $\tilde{\phi}(\mathbf{y}_s)$ cannot be represented exactly with $l$ binary qubits. The peak occurs at the nearest integer to $2^l\tilde{\phi}(\mathbf{y}_s)$, taken modulo $2^l$:
\begin{equation}
j_s=\lfloor 2^l\tilde{\phi}(\mathbf{y}_s)\rceil\bmod 2^l.
\end{equation}

\begin{figure*}[tp]
    \includegraphics[width = 1\linewidth,trim={5cm 0cm 2cm 0cm},clip]{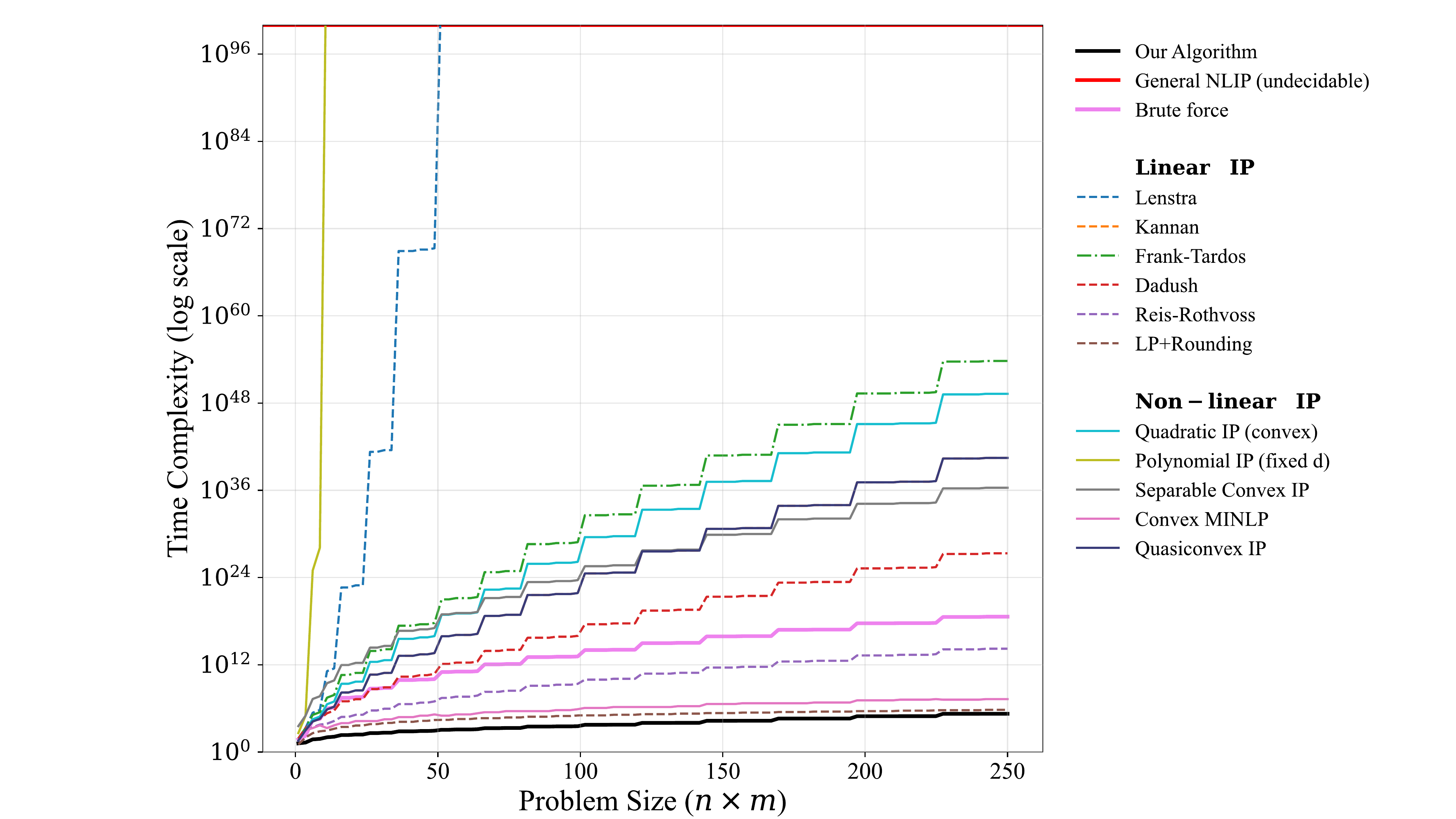}
    \caption{Comparative analysis of integer programming algorithms. Time complexity scaling with problem size ($n \times m$) with $d=5$, $V=10$, $\Delta = 100$, $k=3$, $c=2$ and $\epsilon_{\text{QPE}} = 0.1$.}
\label{fig:combined}
\end{figure*}
\section{Determining the convexity for a non-linear IP}
\label{convex}
The convexity of the nonlinear sample problem given by Eq.~\eqref{ExP1} is analyzed by examining its Hessian matrices and eigenvalues. Here, convexity refers to the constraint functions on the continuous relaxation of the variable domain. A twice continuously differentiable function is convex on a convex open domain if and only if its Hessian is positive semidefinite throughout that domain.

\bigskip

\noindent Constraint 1: $C_1(\bm{x}) = x_1 + x_2^2 x_3 + x_3 - 1 < 0$,
with the non-linear part:
\begin{equation}
 h_1(x_2, x_3) = x_2^2 x_3 + x_3.
\end{equation}
Partial derivatives of \(h_1\) is calculated below,
\begin{equation}
\frac{\partial^2 h_1}{\partial x_2^2} = 2 x_3, \quad
\frac{\partial^2 h_1}{\partial x_2 \partial x_3} = 2 x_2, \quad
\frac{\partial^2 h_1}{\partial x_3^2} = 0,
\end{equation}

to give the Hessian matrix as,
\begin{equation}
H(h_1) =
\begin{bmatrix}
2 x_3 & 2 x_2 \\
2 x_2 & 0
\end{bmatrix},
\end{equation}
and the eigenvalues of \(H(h_1)\) are given by:
\begin{equation}
\lambda_{1,\pm} = x_3 \pm \sqrt{x_3^2 + 4 x_2^2},
\end{equation}
where $\sqrt{x_3^2+4x_2^2}>|x_3|$ for $x_2\neq0$, leading to one positive eigenvalue and one negative eigenvalue. In general,
\begin{equation}
\lambda_{1-} = x_3 - \sqrt{x_3^2 + 4 x_2^2} \leq 0.
\end{equation}
Thus, $H(h_1)$ is indefinite for $x_2\neq0$, and $h_1$ is non-convex on the continuous domain. At $x_2=0$ and $x_3\geq0$, its eigenvalues are $0$ and $2x_3$.

\bigskip

\noindent Similarly for other constraints:
\begin{align}
    C_2(\bm{x}) &= 3 x_3^2 x_4 + x_2 - 2 < 0 \nonumber\\ &h_2(x_3, x_4) = 3 x_3^2 x_4 \nonumber\\ \nonumber\\
    C_3(\bm{x}) &= x_1 x_5 + x_4 - 1 < 0 \nonumber\\
    &h_3(x_1, x_5) = x_1 x_5 \nonumber\\\nonumber\\
    C_4(\bm{x}) = &2 x_1 + 2 x_1^2 x_3 + x_4^3 - 2 < 0\nonumber\\
    h_4(&x_1, x_3, x_4) = 2 x_1^2 x_3 + x_4^3.
\end{align}
The corresponding Hessian matrices are:
\begin{align}
H(h_2) =
&\begin{bmatrix}
6 x_4 & 6 x_3 \\
6 x_3 & 0
\end{bmatrix},\quad
H(h_3) =
\begin{bmatrix}
0 & 1 \\
1 & 0
\end{bmatrix}, \nonumber\\
H(h_4) &=
\begin{bmatrix}
4 x_3 & 4 x_1 & 0 \\
4 x_1 & 0 & 0 \\
0 & 0 & 6 x_4
\end{bmatrix}.
\end{align}
and the eigenvalues are:
\begin{align}
    \lambda_{2,\pm} & = 3 x_4 \pm 3 \sqrt{x_4^2 + 4 x_3^2}, \quad \lambda_{3,\pm} = \pm 1, \nonumber\\
    \lambda_{4,\pm} &= 2 x_3 \pm 2 \sqrt{x_3^2 + 4 x_1^2}, \quad \lambda_{4,3}=6x_4.
\end{align}

The Hessian of $h_2$ is indefinite for $x_3\neq0$, that of $h_3$ is indefinite everywhere, and that of $h_4$ is indefinite for $x_1\neq0$. Each nonlinear constraint function $h_i$ therefore has an indefinite Hessian at points in the continuous domain and is \textbf{non-convex}.

Figure~\ref{fig:combined} compares the classical algorithms for linear and non-linear IP with the quantum algorithm for the parameters described in the caption. The algorithms solving non-linear problems apply to convex IP.

\bibliographystyle{unsrt}
\bibliography{bib.bib}

\end{document}